\newcommand{\ignore}[1]{}
\ignore{This is Version 26}

\documentclass[11pt]{article}
\usepackage[xdvi]{epsfig}

\newcommand{\smfrac}[2]{ {\scriptstyle \frac{#1}{#2} } }
\newcommand{\half}{ {\scriptstyle \frac{1}{2} } }

\newcommand\be{\begin{equation}}
\newcommand\ee{\end{equation}}
\newcommand\bea{\begin{eqnarray}}
\newcommand\eea{\end{eqnarray}}

\newcommand{\text}[1]{\qquad \mbox{#1} \qquad}

\def\tr{{\rm Tr}}

\title{Glueball Spectrum for QCD from $AdS$ Supergravity Duality\thanks{
This work was supported in part by the Department of Energy under
Contracts No. DE-FG02-91ER40676 and No. DE-FG02-91ER40688}}

\author{Richard  C. Brower
\\ Physics Department\\
Boston University\\
Boston, MA 02215, USA \\
\and
Samir D. Mathur\\
Department of Physics\\
Ohio State University \\
Columbus OH 43210, USA \\
\and
Chung-I Tan \\
Physics Department \\
Brown University\\
Providence, RI 02912, USA}

\begin{document}

\maketitle

\begin{abstract}
     We present the  analysis of the  complete glueball
     spectrum for the $AdS^7$
     black hole supergravity dual of $QCD_4$ in strong coupling limit:
     $g^2 N \rightarrow \infty$.  The bosonic fields in the supergravity
     multiplet lead to 6 independent wave equations contributing to
     glueball states with $J^{PC} = 2^{++},1^{+-}, 1^{--}$, $0^{++}$ and
     $0^{-+}$.  We study the spectral splitting and degeneracy patterns
     for both $QCD_4$ and $QCD_3$.  Despite the expected limitations of
     a leading order strong coupling approximation, the pattern of
     spins, parities and mass inequalities bare a striking resemblance
     to the known $QCD_4$ glueball spectrum as determined by lattice
     simulations at weak coupling.
\end{abstract}

\newpage \section{ Introduction}

The Maldacena duality conjecture~\cite{maldacena}~ and its further
extensions~\cite{wittenO,gkp}~ state that there is an exact
equivalence between large N conformal field theories in d-dimensions
and string theory in $\bf {AdS^{d+1} \times M}$.  Subsequently
Witten~\cite{wittenT} suggested how to break explicitly the conformal
(and SUSY) symmetries to arrive at a dual gravity description for
$SU(N)$ quarkless $QCD_4$.  Thus we may have at last a definite
proposal for the long sought ``QCD string''. As anticipated by
't Hooft the dual correspondence for the large $1/N$ expansion for
$SU(N)$ Yang-Mills theory is the perturbative expansion of string
theory. Still this theory is difficult to formulate, let alone
solve. At present explicit calculations also require taking the strong
coupling limit, $g^2N \rightarrow \infty$, where the string tension
goes to infinity ($\alpha' \rightarrow 0$) and the dual theory is
classical gravity.

In this paper we complete the study of the glueball spectrum for the
strong coupling dual description of $QCD_4$. For comparison the
analogous spectrum calculation is presented for $QCD_3$, which shows a
very similar pattern, which in qualitative
terms can be traced to  the underlying  flat space T duality between type
IIA and IIB string theories which in turn are  the AdS  duals to $QCD_4$ and
$QCD_3$ respectively. The goal is to learn more about the AdS/Yang-Mills
correspondence by comparing the AdS strong coupling spectrum with the
rather well determined glueball spectrum~\cite{MP} in lattice gauge
theory.  Of course, the strong coupling expansion at best can provide
a rough guide to the underlying physics.  Nonetheless the
correspondence to the continuum (i.e. weak coupling) limit of the
lattice spectrum is surprisingly good.  This comparison may prove
useful to provide support for the conjectured Maldacena duality and to
give specific information on the major strong coupling artifacts that
must be removed as one approaches universality at the ultraviolet
fixed point.

In addition to new spectral calculations, we summarize the earlier
work by many authors~\cite{csaki,jev,several,oz,bmt1,bmt_others,cm}.
In particular, we extend our earlier paper~\cite{bmt1} on the tensor
glueball for $QCD_3$ on an $AdS^5$ black hole background to the
physically relevant case of $QCD_4$.   For $QCD_3$,
we found that  the tensor spectrum ($2^{++}$)
was degenerate~\cite{Freedman} with dilaton($0^{++}$)  and the axion 
($0^{+-}$).
However, the mass gap  set by a lower scalar  glueball obeys the  inequality,
\be
m(0^{++}) < m(2^{++}) \; .
\label{eq:ineq}
\ee
This scenario is repeated for $QCD_4$. The dilaton mode ($0^{++}$)
coupling to $Tr[F^2]$ remains degenerate with the tensor
($2^{++}$), but the axion ($0^{+-}$) is heavier, consistent with
lattice results.  Due to the trace anomaly, the lowest mass scalar
($0^{++}$) couples to the energy density $T_{00}$ and again it obeys this
inequality above, Eq.~(\ref{eq:ineq}).  As in our earlier work, the goal is
to see the details of the spin structure of the lowest glueball
states, which we believe is most sensitive to the underlying gauge
theory.  We find that our analysis combined with all the earlier
results can give a systematic and complete strong coupling glueball
spectrum.

It is useful to end this introduction with a rough overview of
our results and the organization of the paper.

The geometrical  construction for $QCD_4$ is roughly as follows.  One
starts with 11 dimensional M theory  on ${\bf AdS^7 \times S^4}$.
The seven dimensional $AdS^7$ we take to have a radial co-ordinate $r$
and Euclidean space-time co-ordinates $x_1, x_2, x_3, x_4, x_5$ and
$x_{11}$. The ``eleventh'' dimension is taken to be compact, reducing
the theory to type IIA string theory.  Matter at the center of this
space ($r=0$) consists of N D4-branes (or NS 5-branes wrapping $S^1$ in
the 11th coordinate) with  world volume co-ordinates $x_1
\cdots x_5$.  The 5-d Yang-Mills CFT ``living on'' the brane is
dimensionally reduced to $QCD_4$ by raising the ``temperature",
$\beta^{-1}$, in a direction $x_5= \tau$, parallel to the brane. The
new metric is an $AdS^7$ black hole with $x_{11}$ compact .  Compact
directions on $S^4$ will be denoted by $x_\alpha$, $\alpha=7,8,9,10.$

\begin{figure}[h]
\mbox{\epsfxsize=65mm\epsfysize= 120mm\epsfbox{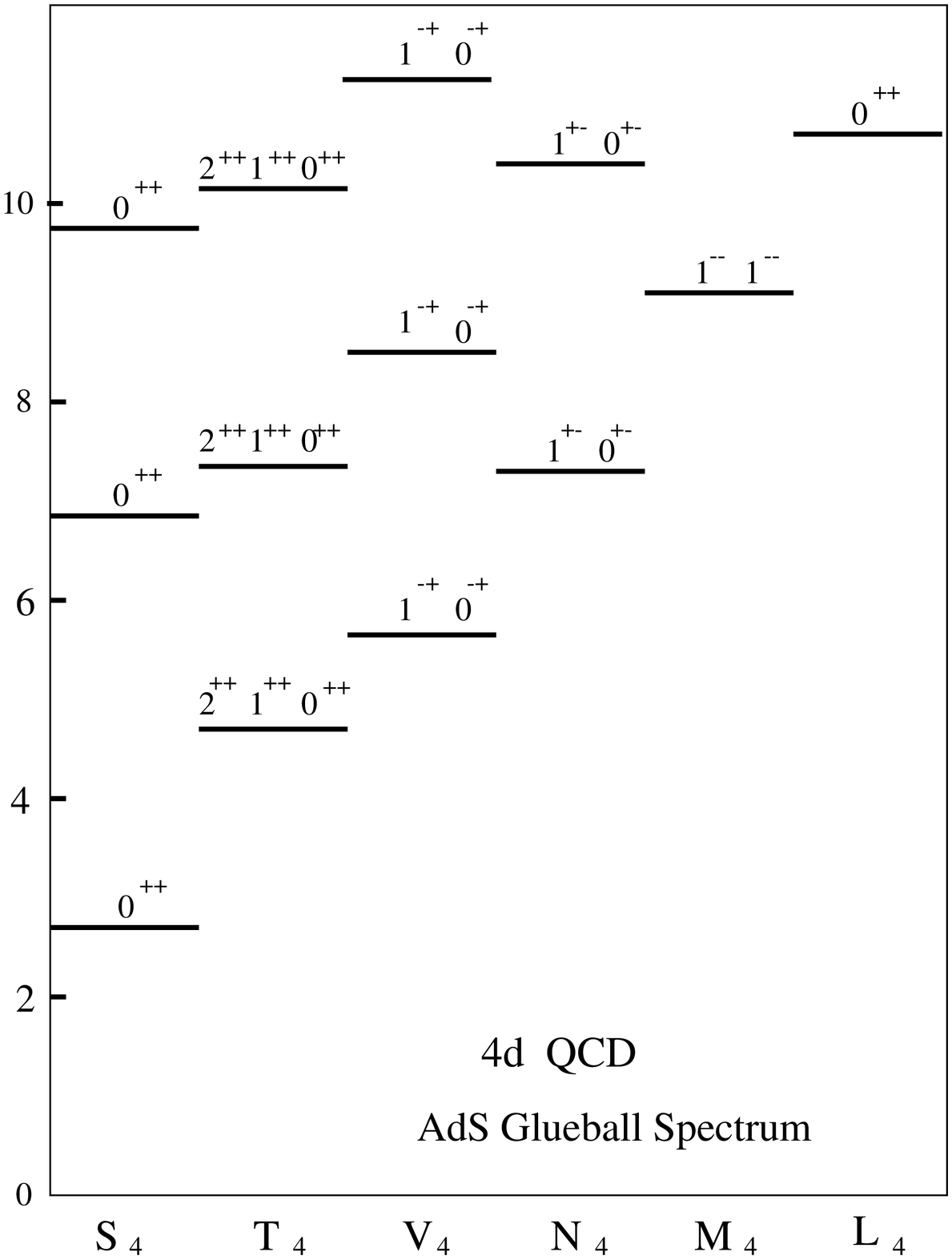} \rule{5mm}{0mm}
        \epsfxsize=70mm\epsfysize= 120mm\epsfbox{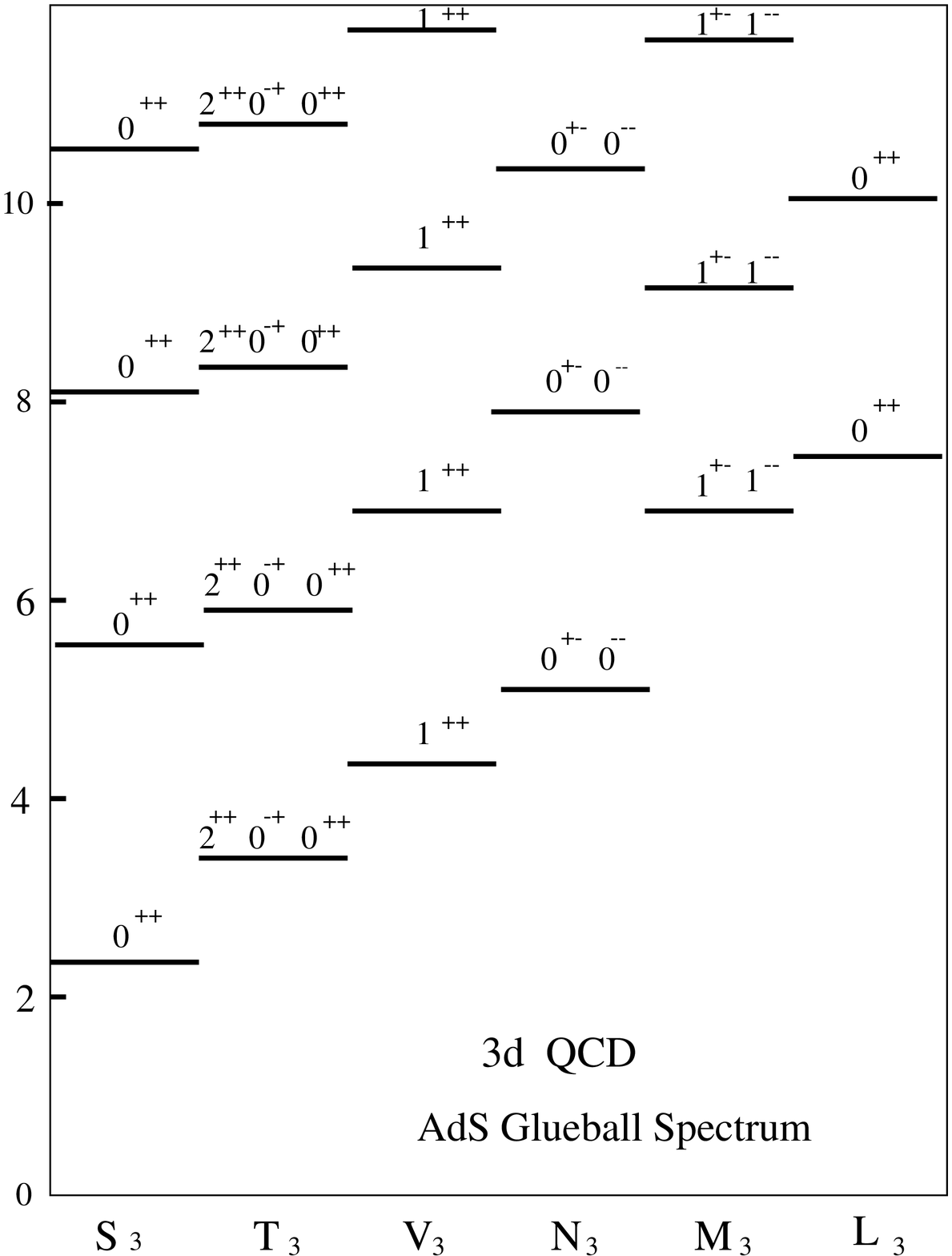}}
        \caption{The AdS glueball spectrum for $QCD_4$ (left) and
        $QCD_3$ (right) with mass eigenvalues, $m_n$, plotted for each of
        the six equations labeled on the horizontal axis.}
        \label{fig:summary}
\end{figure}

The strong coupling glueball calculation consists of finding the
normal modes for the bosonic components of the supergraviton multiplet
in the ${\bf AdS^7 \times S^4}$ black hole background.  We are only
interested in excitations that lie in the superselection sector for
$QCD_4$. So we can ignore modes for all non-trivial harmonics in $S^5$
that carry a non-zero R charge and all Kaluza-Klein (KK) modes in the
two $S^1$ circles with U(1) KK charges.  Imposing these restriction
and exploiting symmetries of the background metric reduce the problem
to {\bf six} independent wave equations, referred to as $S_4,T_4,V_4,
N_4,M_4$ and $L_4$ in the text. In Fig.~\ref{fig:summary} (left side),
we plot the low mass states for each equation, labeling the quantum
numbers for each level.
We can identify the modes with the bosonic components of the zero mass
sector of type IIA string theory: the graviton ($G$), the dilaton
($\phi$), the NS-NS 2-form ($B$) and RR 1- \& 3-forms ($C_{(1)}$,
$C_{(3)}$). The spin degeneracy of the spectrum is due to a spurious
$O(4)$ symmetry of the strong coupling approximation that combines the
11th and three spatial co-ordinates. However, as we explain in the text,
all extra states not observed in the lattice data for the glueball
spectrum, actually carry a discrete ``$\tau$-parity'' that places them, like
KK modes, outside the QCD superselection sector.

We have included in Fig.~\ref{fig:summary} (right side) the full
spectrum for $QCD_3$. The entire analysis is very similar. Starting
with type IIB strings in ${\bf AdS^5 \times S^5}$ which is dual to ${\cal
N}=4$ SUSY Yang-Mills, one introduces a compact thermal circle forming
an $AdS^5$ black hole.  Again there are {\bf six} independent wave
equations, labeled by $S_3,T_3,V_3, N_3,M_3$ and $L_3$.  They
correspond to fluctuations for type IIB fields: the graviton
($G$), the dilaton ($\phi$), the NS-NS 2-form ($B$) as before and RR
0- \& 2-forms ($C_{(0)}$, $C_{(2)}$). For both cases we also include
volume fluctuations in the compact sphere $S^4$ and $S^5$ for type IIA
and IIB respectively.

In Sec. 2, we give general arguments for the spin and degeneracy of
glueball states for $QCD_4$ followed in Sec. 3 by the analysis for
$QCD_3$. For each, we give the resultant six wave equations and
numerical values for the first ten levels. (Derivations for these
equations are explained further in Appendix A.) For all but the lowest
eigenvalue, the glueball masses, $m_n$, are well approximated by the
WKB expansions ~\cite{bmt1,minahan}: $m^2_n = \mu^2 ( n^2 + \delta \; n +
\gamma)$.  For the first level ($n = 0$), we provide a simple but
reasonably tight variational upper bound~\cite{bmt1}. (See Appendix B
for details.)

In Sec. 4, we give the parity and charge conjugation
quantum numbers of the glueball states using the Born-Infeld action to
determine the quantum numbers of the couplings between gravity fields
and gauge fields. The striking similarity between the $QCD_4$ and
the $QCD_3$ spectra (see Fig.~\ref{fig:summary}) can be  understood
qualitatively  in terms of T-duality, which relates $D4$ branes in IIA to $D3$
branes in IIB.

In Sec. 5, we compare the AdS strong coupling
spectrum with the well determined levels from lattice $QCD_4$ and
remark on the relationship to the constituent gluon picture. No extra
states are present in the $AdS$ spectrum that couple to QCD operators,
although the absence of the low mass $2^{-+}$ state is noted. We also
show how the strong coupling expansion for the Pomeron intercept may
be used to provide an estimate of the coupling at the crossover
between the strong  and weak coupling regimes.

\section{Glueball Spectrum for $QCD_4$}

To approach $QCD_4$ one begins with M theory on ${\bf AdS^7 \times S^4}$.
We compactify the ``eleventh'' dimension (on a circle of radius $R_1$)
to reduce the theory to type IIA string theory and then following the
suggestion of Witten raise the ``temperature'', $\beta^{-1}$, with a
second compact radius $R_2$ in a direction $\tau$, with $\beta = 2\pi
R_2$. On the second ``thermal'' circle, the fermionic modes have
anti-periodic boundary conditions breaking conformal and all SUSY
symmetries.  This lifts the fermionic masses and also the scalar
masses, through quantum corrections. The 't Hooft coupling is $g^2 N
=2\pi g_s N l_s/R_2$, in terms of the closed string coupling, $g_s$
and the string length, $l_s$ .  Therefore, in the scaling limit, $g^2
N \rightarrow 0$, if all goes as conjectured, there should be a fixed
point mapping type IIA string theory onto $SU(N)$ pure Yang-Mills
theory.

We consider the strong coupling limit at large N, where the string
theory becomes classical gravity in the $AdS^7$ black hole metric,
\be
ds^2 = (r^2-{1\over r^4}) d\tau^2 + r^2 \sum_{i=1,2,3,4,11} d x_i^2 +
(r^2-{1\over r^4})^{-1} d r^2 +{1\over 4} d\Omega_4^2 \; ,
\ee
with radius of curvature, $R^3_{AdS} = 8 \pi g_s N l_s^3$.  We
have removed all dimensionful parameters in the metric by a normalization
setting $R_{AdS}= 1$ and $\beta=2\pi/3$.

\subsection{Spin and Degeneracy of Glueball States}

In M theory the supergraviton is a single multiplet in 11-d with two
bosonic fields - a graviton, $G_{MN}$, and a 3-form field, $A_{MNL}$, as
designated in Table~\ref{tab:classIIA}. After restricting all indices
and co-ordinate dependence to $AdS^7$, we have a graviton, $G_{\mu
\nu}$, a dilaton $\phi$, and an NS-NS tensor field $B_{\mu \nu}$.
In addition there are two RR fields, a one-form $C_{\mu}$ and
a three-form $C_{\mu \nu\lambda}$.  Furthermore, we will also consider
the scalar modes coming from ``volume" fluctuations for $S^4$. The
relationship between M theory and IIA string theory nomenclature,
after restricting to the $AdS^7$ subspace, is presented in
Table~\ref{tab:classIIA}.  The table gives the $J^{PC}$ quantum
numbers for all glueball states. The pattern of degeneracy (explained
below) is indicated by the rows ending with the lowest eigenvalue for
each of the six wave equations, Eq. (\ref{eq:4dwe}): $T_4, V_4, $ etc.

The task is to find all the quadratic fluctuations in the $AdS^7$ black
hole background that might survive for $QCD_4$ in the scaling (weak
coupling) limit, ignoring any Kaluza-Klein mode in compact manifolds
(compactified $S^1$ for $x_{11}$, for $\tau$ and the spheres
$S^4$). They are charge states in their own superselection sector that
are clearly absent in the putative target theory. Additional
``spurious'' states will be discussed in Sec. 4 where we treat
discrete symmetries.

\begin{center}
\begin{table}[h]
\begin{tabular*}{150mm}{@{\extracolsep\fill}||c|c|c|c| |c|c|c|}
\hline
%
%
%
\multicolumn{4}{||c||}{States from 11-d $G_{M N}$}& \multicolumn{3}{c||}{States
from 11-d $A_{M N L}$}   \\
\hline\hline
$\;\;\; G_{\mu\nu}\;\;\; $          & $G_{\mu,11} $       & $G_{11,11}$  &
$m_0$ (Eq.) &
$A_{\mu\nu,11}$         & $A_{\mu\nu\rho}$ &  $m_0$ (Eq.)                  \\
\hline
$G_{ij}$              & $C_i$      & $\phi$   &                  &
$B_{ij}$   & $C_{123}$     &    \\
$2^{++}$              &  $1_{(-)}^{++}$     & $0^{++}$    & 4.7007
($T_4$)&
$1^{+-}$               & $0_{(-)}^{+-}$      &   7.3059 ($N_4$)
\\
\hline
$G_{i\tau}$           & $C_{\tau}$   &              &                &
$B_{i\tau }$   & $C_{ij\tau}$  &         \\
$1_{(-)}^{-+}$              & $0^{-+}$      &               &  5.6555 ($V_4$) &
$1_{(-)}^{--}$               & $1^{--}$      &     9.1129 ($M_4$)            \\
\hline
$G_{\tau\tau}$       &               &               &                 &
\multicolumn{2}{c|}{$G^\alpha_\alpha $ State} &  \\
$0^{++}$      &                       &                 & 2.7034 ($S_4$) &
\multicolumn{2}{c|}{$0^{++}$}           &   10.7239 ( $L_4$ )             \\
\hline
\end{tabular*}
\caption{IIA Classification for $QCD_4$.  Subscripts to $J^{PC}$ designate
$P_\tau = -1$ .
}\label{tab:classIIA}
\end{table}
\end{center}

To count the number of independent fluctuations for a field of given
spin, we adopt the following method. We   imagine  harmonic plane
waves propagating in the AdS radial direction, $r$,
with  Euclidean time, $x_4$. For example metric fluctuation,
\be G_{\mu\nu} = \bar g_{\mu\nu} + h_{\mu\nu}(x), \ee
in the background, $\bar g_{\mu\nu}$, are taken to have the form,
$h_{\mu\nu}(r,x_4)$. There is no dependence on the spatial co-ordinates,
$x_i = (x_1,x_2,x_3, x_{11})$ and the compactified ``temperature"
direction $\tau$.

\subsubsection{Metric fluctuations}

The four dimensional field theory lives on the hypersurface
co-ordinates, $x_1, x_2, x_3, x_4$, (with $x_5 = \tau$ and $x_{11}$ as
the two compactified coordinates.)  A graviton has two polarization
indices. If we were in flat space time, we could go to a gauge where
these indices took values only among $(x_1, x_2, x_3, x_{11}, \tau)$
and not from the set $(r, x_4)$. The polarization tensor should also
be traceless. This leaves $(5\times 6)/2-1=14$ independent
components. In the AdS space time, we can count the number of graviton
modes the same way, though the actual modes that we construct will
have this form of polarization only at $r\rightarrow\infty$; for
finite $r$, other components of the polarization will be constrained
to acquire nonzero values~\cite{bmt1}.

We identify the spin content of these 14 components in two
steps. First note that the background metric is $SO(4)$ symmetric. (It
is flat in the first four of these directions, $\bar g_{11}=\bar
g_{22}=\bar g_{33}=\bar g_{11,11}=r^2$, while it is ``warped'' in the
$\tau$ direction, $\bar g_{\tau\tau}=r^2-1/r^4$). The system
therefore has $SO(4)$ symmetry leading to {\it three distinct
equations} corresponding to 9, 4 and 1 dimensional irreducible representations
under $SO(4)$. In Table~\ref{tab:classIIA}, these are denoted by $T_4$, $V_4$,
and $S_4$ respectively.

These representations lead us to a degenerate spectrum of spins under
the physical $SO(3)$ rotations in $x_1, x_2, x_3$, which we list below:
\begin{itemize}
\item
$9$-dimensional  representation breaks into $5+3+1$ under $S0(3)$,
\bea
G_{ij}:\hskip 100pt h_{ij} - \smfrac{1}{3} \delta_{ij} h_{kk}\ne 0
&\rightarrow&  ~~\mbox{spin-2},\cr
C_i: \hskip110pt   h_{i,11}=h_{11,i}\ne 0 &\rightarrow&  ~~\mbox{spin-1},\cr
\phi:\hskip15pt  h_{11,11}=-3h_{11}=-3h_{22}=-3h_{33}\ne 0 &\rightarrow &
~~\mbox{spin-0},
\eea

\item
$4$-dimensional  representation breaks into $3+1$ under $S0(3)$,
\bea
G_{i\tau}: \hskip110pt  h_{\tau i}=h_{i\tau}\ne 0 &\rightarrow&
~~\mbox{spin-1},\cr
C_{\tau}: \hskip 105pt h_{\tau,11}=h_{0\tau}\ne 0
&\rightarrow&  ~~\mbox{spin-0},
\eea
\item
singlet under $S0(3)$,
\bea
G{\tau\tau}: \hskip15pt  h_{\tau\tau}=-
4h_{11}=-4h_{22}=-4h_{33}=- 4h_{11,11}\ne 0 &\rightarrow& \mbox{spin-0},
\eea
\end{itemize}
where $i,j,k = 1, 2, 3$.

In addition there is a scalar field, $G_{\alpha}^{\alpha}$, coming from the
metric
on the $S^4$ sphere~\cite{oz}, (with $m_{AdS}^2=72$), which is referred to as
$L_4$ in Table~\ref{tab:classIIA}.

\subsubsection{Three-form fields}

The behavior of the three-form field is discussed briefly in the
Appendix, but we recall some of the main features here.  The wave
equation has a topological mass term which results in the equation
being factorized into two first-order equations, yielding upon iteration
two second-order equations. One solution is a massless 3-form field,
which has solutions that are pure gauge for the case when there is no
dependence on the sphere $S^4$, and is thus to be ignored.  The other
field gives a second-order equation with $m_{AdS}^2=36$, but the fact
that we have a first-order equation as the primary equation reduces
the degrees of freedom effectively to those of a massless 3-form
field. Let the propagation directions be again $(x_4,r)$. If we
consider the component with indices $A_{123}$ then we get a specific
nonzero value also for the components $A_{r\tau,11}$ and $A_{4\tau,
11}$. We can count the independent degrees of freedom by looking
only at components that do not have the propagation directions $x_4, r$
among the indices.  Thus we get the fields listed in
Table~\ref{tab:classIIA}.  Reducing the $SO(4)$ states under rotations
in $x_1, x_2, x_3$ yields:
\begin{itemize}
\item
   $4$-dimensional  representation breaks into $3+1$ under $S0(3)$,
\bea
B_{ij}:\hskip100pt A_{ij,11}\ne 0\rightarrow &\mbox{spin-1}, \cr
C_{123}:\hskip100pt A_{ijk}\ne 0\rightarrow
&\mbox{spin-0},
\eea
\item  $6$-dimensional  representation into $3 + 3$ under $S0(3)$,
\bea
B_{i\tau}: \hskip115pt A_{i\tau,11}\ne 0\rightarrow &
\mbox{spin-1}\;,
\cr
C_{ij\tau}:\hskip115pt  A_{ij\tau}\ne 0\rightarrow & \mbox{spin-1} \; .
\eea
\end{itemize}

The field equations for these states  have amplitudes $N_4$ and
$M_4$ as listed in
Table~\ref{tab:classIIA}.

\subsection{Wave Equations and $QCD_4$ Glueball Spectra}

The wave equations for the metric fluctuations for $QCD_4$ have been
obtained in Ref.~\cite{cm} by analyzing the linearized
Einstein equations about the ${\bf AdS^7\times S^4}$ black hole background
which leads to three independent equations, $T_4$, $V_4$ and
$S_4$~\cite{bmt1,cm}. Here we complete the spectral analysis giving
the numerical values for all glueball masses.  Fluctuations $N_4$,
$M_4$ and $L_4$ can be found similarly, leading again to all together six
independent equations for $QCD_4$, expressed in a manifestly hermitian form:
\bea
   &-& {d\over dr} (r^7 -r) {d\over dr} T_4 (r)  - (m^2r^3) T_4(r) =  0 \; ,
\nonumber  \nonumber \\
   &-& {d\over dr} (r^7 -r) {d\over dr}  V_4(r)-  (m^2r^3 - {9\over
r(r^6-1)}) V_4(r) =  0  \; ,\nonumber \\
   &-& {d\over dr} (r^7 -r) {d\over dr} S_4(r) - ( m^2r^3 +
   {432r^5\over (5 r^6-2)^2} ) S_4(r) =  0  \; , \nonumber \\
   &-& {d\over dr} (r^7 -r) {d\over dr}N_4(r) - (m^2r^3 -27r^5 +  {9\over r})
N_4(r)  =  0  \; , \nonumber \\
   &-& {d\over dr} (r^7 -r) {d\over dr} M_4(r)- (m^2r^3 - 27 r^5- {9r^{5}\over
r^6-1})M_4(r)  =  0  \; ,\nonumber \\
   &-& {d\over dr} (r^7 -r) {d\over dr} L_4(r) - (m^2r^3 - 72r^5)
L_4(r)  =  0  \;
.
\label{eq:4dwe}
\eea
We shall provide a more detailed discussion on how these equations can be
obtained in Appendix A, while concentrating here on establishing our
normalization
convention.

Consider first metric perturbations of the form
\be
h_{\mu\nu}= \epsilon_{\mu\nu}(r) e^{ik_4 x_4}\; ,
\ee
with all other fields set to zero. We shall further fix gauge to
$h_{4\mu}=0$, and from the linearized Einstein's equation, we
determine the discrete spectrum with $k_4=im$.  Because of the $SO(4)$
symmetry in $x_1,x_2,x_3,x_{11}$, the system is highly degenerate.  Three
distinct equations for various perturbations can be obtained by the
following procedure.

{\bf Tensor:} There are five independent perturbations which form the
spin-2 representations of $SO(3)$:
\be
h_{ij}=q_{ij} r^2 T_4(r)  e^{-mx_4},
\ee
where $i,j=1,2,3$ and $q_{ij}$ is an arbitrary constant traceless-symmetric
$3\times
3$   matrix.

{\bf Vector:}  Consider perturbations:
\be
h_{i\tau}=h_{\tau i} =q_i
\sqrt{r^6-1} V_4(r) e^{-mx_4},
\ee
where $i=1,2,3$ and $q_i$ is an arbitrary constant 3-vector.
Both equations for $T_4$ and $V_4$ have also been obtained in Ref.~\cite{oz}
by considering the corresponding degenerate scalar modes.

   {\bf Scalar:}  The analogous
scalar perturbation is
\be
h_{\tau\tau} = (r^2-r^{-4}) S_4(r)\; e^{-mx_4} \; .
\ee

{\bf Three-form and Volume Scalar:}
Next we turn to 3-form fields. It is sufficient to consider
\bea
A_{ij,11}=B_{ij} &=&\epsilon_{ijk} q_k r^2 N_4(r) e^{-mx_4}\;,\cr
A_{i\tau,11}=B_{i\tau} &=& q_i\sqrt{r^6-1} M_4(r) e^{-mx_4}\; ,
\eea
   where $q_i$ is again
an arbitrary constant 3-vector.

Note that the metric in the direction $x_{11}$ is the
same as that in the directions $i,j,k$, so the above functions $N_4$,
$M_4$ will also
give the solutions for $C_{ijk}$ and $C_{ij\tau}$ respectively.
(Fluctuations for
$B_{ij}$ have been considered previously in Ref.~\cite{csaki}. )

Lastly, for the volume perturbation,
we consider~\cite{oz}
$$h^{\alpha}_{\alpha}= L_4(r) e^{-m x_4}. $$

\vskip - 5mm
\begin{center}
\begin{table}[th]
\begin{tabular*}{150mm}{@{\extracolsep\fill}|l|r|r|r|r|r|r|}
\hline
Equation:     &$T_4$ &$V_4$  & $S_4$  & $N_4$ & $M_4$       &$L_4$          \\
$J^{PC}$:     & $2^{++}/1^{++}/0^{++}$ & $1^{-+}/0^{-+}$         &
$ 0^{++}$ &  $1^{+-}/0^{+-}$ & $1^{--}/1^{--}$         &$0^{++}$             \\
\hline
n = 0 &     22.097 &     31.985 &      7.308 &
        53.376 &     83.046 &    115.002 \\
\hline
n = 1 &     55.584 &     72.489 &     46.986 &
       109.446 &    143.582 &    189.631 \\
\hline
n = 2 &    102.456 &    126.174 &     94.485 &
       177.231 &    217.399 &    277.282 \\
\hline
n = 3 &    162.722 &    193.287 &    154.981 &
       257.958 &    304.536 &    378.099 \\
\hline
n = 4 &    236.400 &    273.575 &    228.777 &
       351.895 &    405.018 &    492.169 \\
\hline
n = 5 &    323.541 &    368.087 &    315.976 &
       459.131 &    518.059 &    619.547 \\
\hline
n = 6 &    424.195 &    474.268 &    416.666 &
       579.706 &    646.088 &    760.252 \\
\hline
n = 7 &    538.487 &    594.231 &    530.950 &
       713.638 &    786.559 &    914.307 \\
\hline
n = 8 &    666.479 &    729.102 &    658.996 &
       860.939 &    939.557 &   1081.732 \\
\hline
n = 9 &    808.398 &    875.315 &    800.860 &
      1021.613 &   1108.010 &   1262.518 \\
\hline
\end{tabular*}
\caption{The mass spectrum, $m_n^2$, for $QCD_4$ Glueballs }  \label{tab:mass4}
\end{table}
\end{center}

To calculate the discrete spectrum for each of these equations, one must
apply the correct boundary conditions at
$r = 1$ and $r = \infty$. This issue has been discussed in several earlier
papers~\cite{jev,bmt1}. The boundary conditions are found by solving the
indicial
equation. In all cases the appropriate boundary condition~\cite{jev}~ at
$r =1$ is the one without the logarithmic singularity.  At $r =
\infty$ the least singular boundary is required to have a normalizable
eigenstate.  (See Appendix B for a listing of all boundary conditions.)
Matching  boundary conditions from $r=1$ and $r=\infty$ results in a
discrete set of eigenvalues $m^2_n$, where $n$ is the number of zeros
in the wave function inside the interval $r \in (1,\infty)$.  We
solved the eigenvalue equations  by the shooting method, integrating
from $r_1 \simeq 1$ to large $r_\infty \simeq \infty$.
The resultant spectrum is given in Table~\ref{tab:mass4}.

To further  check our results, we  have compared our numerical masses to
the WKB approximations,
\be
    m^2_n \simeq \; \mu_4^2 ( n^2 + \delta \; n +
\gamma) + 0({1\over n})\>,
\ee
where $\mu^2_4= 36 \pi (\Gamma(2/3)/ \Gamma(1/6) )^2$.
For each equation, individual integer constant, $\delta$, was
determined analytical and constant $\gamma$ was fit to the
numerical data in Table~\ref{tab:mass3}.
(See Appendix B).  The fits to
the WKB formula are accurate to better than 0.1 \% for all but the
lowest ($n=0$) mode. For each lowest mode, we have also carried out an
independent simple variational estimate.  We note that in  case,
our numerically calculated value for $m^2_0$ is always close and
respects the variational estimate as an upper bound.
(See Table~\ref{tab:variational}.)

\section{Glueball Spectrum for $QCD_3$}

For $QCD_3$ the construction of the supergravity dual begins with
Maldacena's conjecture for type IIB string theory in a ${\bf AdS^5 \times
S^5}$ background metric. Here the dual theory is conjectured to be the
conformal field theory for ${\cal N} = 4$ SUSY SU(N) Yang-Mills in
4-d.  The $AdS$ curvature is induced by N units of charge on N
coincident D3 branes giving rise to a constant volume 5-form $F_{(5)}$
in the product manifold. The co-ordinates in the $AdS^5$ space, we
label by one ``radial'' co-ordinate $r$ and four space-time co-ordinates,
$x_\mu$, $\mu= 1,2,3,4$, parallel to the D3 branes. The remaining five
co-ordinates in $S^5$ are labeled by $x_\alpha$, $\alpha=6,7,8,9,10$.

Following the suggestion of Witten for obtaining a supergravity dual
to $QCD_3$, we break conformal and SUSY symmetries, by introducing a
compact ``thermal'' co-ordinate $x_4 = \tau$ with anti-periodic
boundary on $S^1$ for the fermionic modes. The resultant metric at
high temperature is an $AdS^5$ black hole,
\be
ds^2 = (r^2-{1\over r^2}) d\tau^2 + r^2 \sum_{i=1,2,3} d x_i^2 +
(r^2-{1\over r^2})^{-1} d r^2 + d\Omega_5^2 \; ,
\ee
with radius of curvature, $R^4_{AdS} = 4\pi g_s N l_s^4$ and 3-d
Yang-Mills coupling, $g^2_3 N = 2g_s N/R$ in terms of the string
coupling $g_s$, string length $l_s$ and compact $S^1$ circumference $\beta
= 2 \pi R$. We have removed all dimensionful parameters from the
metric by adopting a simple normalization with $R_{AdS} = 1$ and
$\beta=\pi$, for the circumference of the thermal circle.  At high
temperature (or equivalently low energies), IIB string theory in this
background is conjectured to equivalent to $QCD_3$.

\subsection{Spin and Degeneracy of Glueball States}

Type IIB string theory at low energy has a supergravity multiplet with
several zero mass bosonic fields: a graviton, $G_{\mu \nu}$, a
dilaton $\phi$, an axion (or zero form RR field) $C$ and
two tensors, the NS-NS and RR fields $B_{\mu \nu}$ and $C_{\mu \nu}$
respectively. In addition there is the 4-form RR field $C_{(4)}$ that is
constrained to have a self-dual field strength, $F_{(5)} = dC_{(4)}$.

Now the task is to find all the quadratic fluctuations in the above
background metric whose eigen-modes correspond to the discrete
glueball spectra for $QCD_3$ at strong coupling. We are only
interested in excitations that lie in the superselection sector for
$QCD_3$. Thus for example we can ignore all non-trivial harmonic in
$S^5$ that carry a non-zero R charge and all Kaluza-Klein (KK) modes in the
$S^1$ thermal circle with a U(1) KK charge. The result of these
considerations, discussed in detail below are summarized in
Table~\ref{tab:classIIB}.
\begin{table}[h]
\begin{center}
\begin{tabular*}{150mm}{@{\extracolsep\fill}||c|c|c|c| |c|c|c|}
\hline
%
%
%
%
\hline
$\;\;\; G_{\mu\nu}\;\;\;$ & \multicolumn{2}{c|}{$e^{\textstyle -\phi} + i C$
states}   &
    $m_0$ (Eq.)         &
$B_{\mu\nu}$    & $C_{\mu\nu}$  &  $m_0$ (Eq.)                  \\
\hline\hline
$G_{ij}$     &    $C$                      &   $\phi$          &      &
$B_{ij}$         & $C_{ij}$      &                    \\
$2^{++}$      &  $0^{-+}$   &     $0^{++}$        & 3.4041 ($T_3 $) &
$0^{+-}$        & $0^{--}$     &  5.1085  ($N_3$)     \\
\hline
   $G_{i\tau}$   &                       &                  &               &
$B_{i\tau}$      & $C_{i\tau}$   &   \\
   $1^{++}$     &                       &                 & 4.3217  ($V_3$) &
$1^{+-}$        & $1^{--}$      & 6.6537   ($M_3$)  \\
\hline
$G_{\tau\tau}$&                          &                  &    &
\multicolumn{2}{c|}{$G^\alpha_\alpha $ State} &        \\
$0^{++}$      &                       &                 & 2.3361 ($S_3$)  &
\multicolumn{2}{c|}{$0^{++}$}           & 7.4116  ($L_3$) \\
\hline
\end{tabular*}
\caption{IIB Classification for  $QCD_3$}\label{tab:classIIB}
\end{center}
\end{table}

To count the number of independent fluctuations for a field of given
spin, we again imagine  harmonic plane
waves  propagating in the AdS radial direction, $r$,
with Euclidean time, $x_3$.  For example,
the metric fluctuations in $AdS^5$
\be G_{\mu\nu} = \bar g_{\mu\nu} + h_{\mu\nu}(x), \ee
in the fixed background $\bar g_{\mu\nu}$ are taken to be of the
form $h_{\mu\nu} (r,x_3)$.  There is no dependence on the spatial
co-ordinates, $x_i = (x_1,x_2)$, or the compactified ``temperature''
direction, $\tau$.

\subsubsection{Metric fluctuations}

A graviton has two polarization indices. If we were in flat space
time, we could go to a gauge where these indices took values only
among $(x_1, x_2, \tau)$ and not from the set $(r, x_3)$. The
polarization tensor should also be traceless. This leaves $(3\times
4)/2-1=5$ independent components. In the AdS space time, we can count
the number of graviton modes the same way, though the actual modes
that we construct will have this form of polarization only at
$r\rightarrow\infty$; for finite $r$, other components of the
polarization will be constrained to acquire nonzero
values~\cite{bmt1}.

Therefore, a set of {\it five independent} polarization tensors can be
characterized by the following non-vanishing components at { $r\rightarrow
\infty$}. In the $AdS^5$ black hole background, $\tau$ is compact, so the
rotations group is $SO(2)$ in $(x_1, x_2)$ and the five states are in 3
irreducible representations: A spin-2 doublet (helicities $\pm 2$), spanned by
\be
G_{ij}: \hskip60pt h_{12}=h_{21}\ne 0,  \quad\quad {\rm and}  \quad\quad
h_{11}=-h_{22}\ne 0 \; ,
\ee
a spin-1 doublet  (helicities $\pm 1$), spanned by
\be
G_{i\tau}: \hskip60pt h_{\tau 1}= h_{1\tau}\ne 0, \quad\quad {\rm and}
\quad\quad
    h_{\tau 2}=h_{2,\tau}\ne 0
\ee
and  a spin-0 state,
\be
G_{\tau\tau}: \hskip80pt h_{\tau\tau} =-2h_{11}=-2h_{22}\ne 0 \; .
\ee
These fluctuations are denoted by $T_3$, $V_3$, and $S_3$ respectively in Table
\ref{tab:classIIB}.

\subsubsection{Two-form fields}

Each 2-form field in $AdS^5$ satisfies a field equation that includes
a topological mass term.  The 2-forms $B_{\mu\nu}$ and $C_{\mu\nu}$
can be combined into one complex 2-form field $\tilde
B_{\mu\nu}=B_{\mu\nu}+iC_{\mu\nu}$.  The field equation for $\tilde
B_{\mu\nu}$ can be factorized into two first order equations, and each
can be iterated leading to a second order equation of the form
$${\rm Max} \> \tilde B_{\mu\nu}+m_{AdS}^2 \tilde B_{\mu\nu}=0 \; ,$$
where ${\rm Max}$ is the Maxwell operator on 2-forms. For modes that
have no dependence on the $S^5$ co-ordinates, one equation is massive, with
$m_{AdS}^2=16$, and the other is massless. It can be shown that the massless
equation has only pure gauge solutions; so they can be ignored.
(See Appendix A for further details.)

For the purpose of counting modes, polarizations for a massless 2-form
in $AdS^5$ can also be restricted to be transverse, with fields
depending only on $(x_3, r)$.  Therefore the polarization tensor is an
antisymmetric 2-tensor in the directions $x_1, x_2,\tau$, leading to 3
independent components. On the other hand, for a general massive
2-form field, longitudinal polarizations are allowed, so that the
polarization is an antisymmetric tensor in the coordinates $x_1, x_2,
\tau, r$, with 6 independent components.

Nevertheless, the number of independent components for our massive
2-form $\tilde B_{\mu\nu}$ is only 3 (complex), as if we are dealing with a
massless case.  This is due to the fact that the second order equation
above stems from a first order equation,
$\epsilon_{\mu\nu}{}^{\sigma\lambda\rho}\partial_{[\sigma}\tilde
B_{\lambda \rho]}+4i\tilde B_{\mu\nu}=0$, relating real and imaginary
parts and leading to additional constraints.  For instance, with fields
depending on $(x_3,r)$, if we start with the polarization tensor having
$B_{12}=- B_{21}\ne 0\>,$ then the first order equation will constrain
us to have specific values for $C_{\tau r}=- C_{r\tau}\ne 0$ and
$C_{3r}=-C_{r3}\ne 0$, while allowing all other components of $B$ and 
$C$ to be zero.
Thus we count as independent fields $B_{12},B_{1\tau}, B_{2 \tau}, 
C_{12}, C_{1\tau}, C_{2\tau}$, i.e., there are three independent
solutions for $B_{\mu\nu}$ and three solutions for $C_{\mu\nu}$.  

The non-vanishing polarizations of the
$B_{\mu\nu}$ tensor can be grouped as:
\bea
B_{ij}: \hskip80pt B_{12}=-B_{21} &\ne& 0 \; ,
\eea
corresponding to spin-0 and
\bea
B_{i\tau}: \hskip80pt B_{1\tau}=-B_{\tau 1}&\ne& 0,\cr
B_{2\tau}=-B_{\tau 2}&\ne& 0.
\eea
corresponding to a spin-1 doublet. In Table~\ref{tab:classIIB}, these
fluctuations are denoted by $N_3$ and $M_3$ respectively. They are
degenerate with ones for the  R-R 2-form $C_{\mu\nu}$.

\subsubsection{Scalar fields}

In general, for each scalar field, there is a unique field equation
with the plane wave dependence which we have been considering. There
are three such scalar modes: The volume fluctuations
$G^{\alpha}_{\alpha}$ in $S^5$, (with $m_{AdS}^2=32$), denoted by
$L_3$~\cite{csaki}, and fluctuations for the dilaton $\phi$ and the
axion $C$. However, as we have shown in an earlier paper~\cite{bmt1},
the latter two spectra are degenerate with the $2^{++}$ tensor
fluctuations.  Therefore, separate equations are not
required~\cite{bmt1,bmt_others,cm}.

\subsection{Wave Equations and $QCD_3$ Glueball Spectrum}

Metric fluctuations for $QCD_3$ have been obtained previously by
analyzing the linearized Einstein equations about the ${\bf AdS^5\times
S^5}$ black hole background which leads to three independent equations,
$T_3$, $V_3$ and $S_3$~\cite{bmt1,bmt_others,cm}.  Fluctuations
$N_3$, $M_3$ and
$L_3$ can be found similarly, leading to all together six independent
equations for $QCD_3$.   From the equation of
motion, we determine the
discrete spectrum with $k_3=im$. (See Appendix A for details.)
   The full set of independent
equations are:
\bea
   &-& {d\over dr } (r^5 - r) {d\over dr}  T_3(r)- (m^2 r )  T_3(r) = 0\;,
\nonumber \\
   &-& {d\over dr } (r^5 - r)   {d\over d r}  V_3(r) -  (m^2 r -{4\over
r( r^4-1)}
) V_3(r) =  0 \; . \nonumber \\
   &-& {d\over dr } (r^5 - r) {d\over dr}   S_3(r) -
   (m^2r +  {64r^2\over(3r^4-1)^2} )   S_3(r)  =  0 \; ,    \nonumber \\
   &-& {d\over dr } (r^5 - r) {d\over dr}   N_3(r) - (m^2r - 12r^3 +
{4\over r}) N_3(r) =  0    \; ,\nonumber \\
   &-& {d\over dr } (r^5 - r) {d\over dr}  M_3(r)- (m^2r - 12r^3
   -{4r^3\over r^4-1})  M_3(r)  = 0  \; , \nonumber \\
   &-& {d\over dr } (r^5 - r) {d\over dr}  L_3(r) - (m^2r -
32r^3)L_3(r) =  0 \; .
\eea

Each equation can be expressed in a variety of forms, depending on the
choice of normalization.  The following choices have been made so that
each equation takes on a manifestly hermitian form:

{\bf Tensor:}
\be
h_{ij}=q_{ij} r^2 T_3(r)  e^{-mx_3},
\ee
where $i,j=1,2$, with $q_{ij}$  an arbitrary constant traceless-symmetric
$2\times
2$   matrix.

{\bf Vector:}
\be
h_{i\tau}= q_i \sqrt{r^4-1} V_3(r) e^{-mx_3},
\ee
where $q_i$ is a constant 2-vector.

   {\bf Scalar:} This  case has been treated carefully
using several different gauge
choices~\cite{bmt1,cm}. Here we adopt the form suggested
by Constable and Meyer~\cite{cm}
with
\be
h_{\tau\tau} = (r^2-r^{-2}) S_3(r)
e^{-mx_3}.
\ee

{\bf Two-forms and volume scalar:} For $B_{12}$, consider
perturbations of the form~\cite{csaki}
\be
B_{12}= r^2 N_{3}(r) e^{ik_3 x_3},
\ee
with $B_{1\tau}=B_{2\tau}=0$. Alternatively,
we consider
\be
B_{i\tau} = q_i \sqrt{r^4-1} M_3(r) e^{-mx_3},
\ee
with $B_{12}=0$, where $q_i$ is an arbitrary constant 2-vector.
Lastly, for the volume perturbation,  we consider
$$h^{\alpha}_{\alpha}= L_3(r) e^{-m x_3}. $$

To calculate the discrete spectrum for each of these equations one
must again apply the correct boundary conditions at $r = 1$ and $r =
\infty$ as mentioned in the case of  $QCD_4$ earlier.
(See Appendix B for a listing of boundary conditions.)  The resultant
spectrum is given in Table~\ref{tab:mass3}.

\begin{table}
\begin{center}
\begin{tabular*}{150mm}{@{\extracolsep\fill}|l|r|r|r|r|r|r|}
\hline
Equation:     &$T_3$ &$V_3$  & $S_3$  & $N_3$ & $M_3$       &$L_3$          \\
$J^{PC}$:     & $2^{++}/0^{-+}/0^{++}$ & $1^{++}$         & $0^{++}$ &
      $0^{+-}/0^{--}$ & $1^{+-}/1^{--}$         & $0^{++}$             \\
\hline
n = 0 &     11.588 &     18.677 &      5.457 &
        26.097 &     44.272 &     54.932 \\
\hline
n = 1 &     34.527 &     47.499 &     30.442 &
        61.159 &     84.318 &    100.628 \\
\hline
n = 2 &     68.975 &     87.720 &     65.123 &
       107.308 &    135.932 &    157.933 \\
\hline
n = 3 &    114.910 &    139.436 &    111.141 &
       164.829 &    199.073 &    226.773 \\
\hline
n = 4 &    172.331 &    202.623 &    168.601 &
       233.791 &    273.717 &    307.123 \\
\hline
n = 5 &    241.237 &    277.283 &    237.528 &
       314.215 &    359.861 &    398.971 \\
\hline
n = 6 &    321.627 &    362.779 &    317.931 &
       406.112 &    457.507 &    502.311 \\
\hline
n = 7 &    413.501 &    461.121 &    409.815 &
       509.486 &    566.631 &    617.140 \\
\hline
n = 8 &    516.860 &    568.462 &    513.180 &
       624.341 &    687.279 &    743.457 \\
\hline
n = 9 &    631.703 &    689.156 &    628.028 &
       750.677 &    819.329 &    881.260 \\
\hline
\end{tabular*}
\caption{The mass spectrum, $m_n^2$, for $QCD_3$ Glueballs }  \label{tab:mass3}
\end{center}
\end{table}
Similarly, we  have also compared our numerical $QCD_3$ masses to
the WKB approximations,
\be
    m^2_n \simeq \; \mu_3^2 ( n^2 + \delta \; n +
\gamma) + 0({1\over n})\>,
\ee
where $\mu^2_3= 16 \pi ( \Gamma(3/4)/ \Gamma(1/4) )^2$, with integer
constants, $\delta$, listed in
Table~\ref{tab:WKB}, determined analytically. Again, the constants
$\gamma$ were fits to
the numerical data.  (See Appendix B).

\section{Parity and Charge Conjugation assignments}

Next we determine how the supergravity fields and therefore the
glueballs couple to the boundary gauge theory.  This allows us to
unambiguously assign the correct parity and charge quantum numbers to
the glueball states. For this purpose we consider the effective
Born-Infeld action on the branes.

\subsection{$QCD_4$}

The 4-d gauge theory for $QCD_4$ is obtained by dimensional reduction
from a 5-d gauge theory, which is the low energy dynamics of D4-branes
in 10-dimensional Type IIA string theory.  Although this 10-d theory
may itself be regarded as a dimensional reduction of 11-d M-theory for
membranes, it is sufficient and more convenient to consider the 10-d
theory itself.

Since supergravity fields can be thought of as coupling constants for
gauge theory operators, their quantum numbers can be assigned by the
parity and charge conjugation invariance of the overall action,
(supergravity field times composite operator).  For simplicity let us
consider the  coupling of a supergravity field to just one D4-brane
--- this coupling is given by a Born-Infeld action
plus a Wess-Zumino term,
\be
S=\int d^5x
\det[G_{\mu\nu}+e^{-\phi/2}(B_{\mu\nu}+F_{\mu\nu})]+\int d^4x (C_1
F\wedge F+ C_3\wedge F+ C_5)\; ,
\ee
where $\mu, \nu =  1,2, 3, 4,\tau$. Later we will argue
that our quantum number assignment is correct also for the non-abelian
case of N coincident D branes.  In the 5-d field theory, we have the
space-time world volume co-ordinate $x_1, x_2, x_3,x_4,\tau$ with
$\tau$  compactified on $S^1$. The Euclidean time coordinate we take
to be $x_4$.  After dimensional reduction the physical fields will be
characterized by their representation under the little group $SO(3)$
of rotations  on the spatial co-ordinates $x_i $, $i=1,2,3$, in the 4-d theory.

For the 5-d gauge fields, we define parity by
\bea
P&:& A_i(x_i,x_4,\tau)\rightarrow  -A_i(-x_i,x_4,\tau),\cr
P&:& A_4(x_i,x_4,\tau)\rightarrow \> \>\>\>A_4(-x_i,x_4,\tau),\cr
P&:& A_{\tau}(x_i,x_4,\tau)  \rightarrow \>\>\>\> A_{\tau}(-x_i,x_4,\tau) \; ,
\eea
for $x_i\rightarrow -x_i$, $x_4\rightarrow x_4$, and $\tau\rightarrow
\tau$.  For the Euclidean ${\bf R^5}$ space, this is the only discrete
symmetry.  However after compactification to ${\bf R^4 \times S^1}$, we
can define another parity (not related by a 5-d proper Lorentz
transformation) by inverting the $\tau$ co-ordinate on ${\bf S^1}$. Thus
we define a separate discrete $\tau-$parity transformation $P_{\tau}:$
$\tau \rightarrow -\tau$,
\bea
P_\tau&:& A_i(x_i,x_4,\tau)\rightarrow \> \>\>\>A_i(x_i,x_4,-\tau),\cr
P_\tau&:& A_4(x_i,x_4,\tau)\rightarrow \> \>\>\>A_4(x_i,x_4,-\tau),\cr
P_\tau&:& A_{\tau}(x_i,x_4,\tau) \rightarrow -A_{\tau}(x_i,x_4,-\tau) \; .
\eea
Charge conjugation for a non-abelian gluon field is
\be
C: \half T_a A^a_\mu(x) \rightarrow - \half T^*_a A^a_\mu(x)
\ee
where $T^a$ are the Hermitian generators of the group.  In terms of
matrix fields ($A \equiv\half T_a A^a$), $ C: A_\mu(x) \rightarrow
- A^T_\mu(x).$ This leads to a subtlety. For example consider the
transformation of a trilinear gauge invariant operators,
\be C: Tr[ F_{\mu_1 \nu_1} F_{\mu_2 \nu_2} F_{\mu_3 \nu_3}]
\rightarrow - Tr[ F_{\mu_3 \nu_3} F_{\mu_2 \nu_2} F_{\mu_1 \nu_1} ] \; .
\ee
The order of the fields is reversed. Hence the symmetric products,
$d^{abc} F^a_1 F^b_2 F^c_3$, have $C = -1$ and the antisymmetric products,
$f^{abc} F^a_1 F^b_2 F^c_3$,  $C = +1$. Of course using a single
brane, we can only find symmetric products. For reasons explained
further in Sec. 5, we will only encounter symmetric traces
over polynomials in F, designate by $Sym~\tr[F_{\mu\nu} \cdots]$. Even
polynomials have $C = +1$ and odd polynomials $C = -1$.

\subsubsection{Graviton couplings}

Expanding the Born-Infeld action, we
can now read off the $J^{PC} (P_\tau)$ assignments:

>From the coupling,
$G_{\mu\nu}T^{\mu\nu}\sim G_{\mu\nu}\tr(F_{\mu\lambda}F^\lambda_\nu )
    + \cdots  \; ,$
we obtain
\be
G_{ij}~\rightarrow ~2^{++}\;\; (P_\tau = +) ,\quad\quad
G_{i\tau}~\rightarrow ~1^{-+}\;\; (P_\tau = +) \quad\quad
G_{\tau\tau}\rightarrow ~0^{++}\;\; (P_\tau = +) \; . \ee
Under compactification of 11-d supergravity theory, $G_{\mu ,11}$
becomes the Ramond-Ramond 1-form $C_{\mu}$, which couples as
$\sim\epsilon^{\mu\nu\lambda\kappa\eta} C_\mu\;
Sym~\tr[F_{\nu\lambda}F_{\kappa\eta} W]\;,$ where $W$ is an
an even power of fields $F$.
Consequently, the coupling,
$ \epsilon^{ijk}C_{i}\;Sym~\tr[F_{\tau j}F_{k4}W]+\cdots$
leads to
\be
C_i ~\rightarrow ~1^{++}\;\; (P_\tau = -) \;.\ee
Similarly, $ \epsilon^{ijk}C_{\tau }\tr(F_{ij}F_{4k}W)+\cdots$
gives
\be
C_\tau ~\rightarrow ~0^{-+}\;\; (P_\tau = +)\; ,
\ee
and $G_{11, 11}$ leads to the dilaton $\phi$ with coupling
$\phi \tr{F^2} \; ,$
\be
\phi ~\rightarrow ~0^{++}\;\; (P_\tau = +) .
\ee

\subsubsection{Two-form, three-form fields, and volume scalar}
Consider first the NS-NS 2-form field $B_{\mu\nu}$.  This field in the
Type IIA theory arises from the 3-form field of the 11-d supergravity
theory when the components of the 3-form field are $A_{\mu\nu,{11}}$. For
$U(1)$ gauge theory in
leading order this field couples as $B_{\mu\nu}F^{\mu\nu}$.
More generally in  the $SU(N)$ gauge
theory, we must have
a multi-gluon coupling, $B_{\mu\nu}\; Sym\tr[ F_{\mu\nu} W ]$.
(Again $W$ is an
an even power of fields $F$ and the  trace is symmetrized.)

To determine the parity, assume for the
supergravity modes that we are in a gauge where the indices of the 2-form
do not point along $x_4, x_{11}, r$.
With $i,j=1,2, 3$,
this leads to coupling $B_{ij}\;Sym\tr[F^{ij}W]$ with
\be
B_{ij} ~\rightarrow ~1^{+-}\;\; (P_\tau = +) \; ,
\ee
and coupling $B_{i\tau}\;Sym\tr[F^{i\tau}W]$  with
\be
B_{i\tau} ~\rightarrow ~1^{--}\;\; (P_\tau = -) \; .
\ee

An analogous analysis can also be carried out for 3-form fields,
$C_{ijk}$ and $C_{ij\tau}$.
The coupling, $C_{123}\;Sym\tr[F^{4\tau}W]$
leads to
\be
C_{ijk} ~\rightarrow ~0^{+-}\;\; (P_\tau = -) \; ,
\ee
and coupling  $\epsilon^{ijk}C_{ij\tau}\;Sym\tr[F^{4 k}W]$
leads to
\be
C_{ij\tau} ~\rightarrow ~1^{--}\;\; (P_\tau = +) \; .
\ee

Lastly, the volume scalar couples as $h^\alpha_\alpha \;Sym\tr F^4+\cdots$
giving
\be
h^\alpha_\alpha ~\rightarrow ~0^{++}\;\; (P_\tau = +) \; .
\ee
The complete parity and charge conjugation assignments are given in
Table~\ref{tab:classIIA}.

\subsection{$QCD_3$}

The 3-d gauge theory is obtained by
dimensional reduction from a 4-d gauge theory.  To find the symmetries
of the interactions, we consider  the
Born-Infeld action plus Wess-Zumino term, describing the coupling of a
supergravity field to  a single D3-brane,
$$S=\int d^4x
\det[G_{\mu\nu}+e^{-\phi/2}(B_{\mu\nu}+F_{\mu\nu})]+\int d^4x (C_0
F\wedge F+ C_2\wedge F+ C_4)\> \; ,$$
where $\mu, \nu\ = 1,2, 3, \tau$.  As in the case of $QCD_4$, we will
find the charge conjugation and parity assignments with the help of
the symmetries of the 4-d gauge theory and then take the dimensional
reduction to the 3-d theory after compactification of the coordinate
$\tau$.  The Euclidean time is taken to be $x_3$ and the spatial
co-ordinates, $x_i$, $i=1,2$.

After dimensional reduction the physical fields will be characterized
by their representation under the little group of the space co-ordinates
of the 3-d theory: this is the group $SO(2)$ rotations in the $x_1,
x_2$ plane. However, unlike the case of $QCD_4$, the usual spatial
inversion, $x_i\rightarrow -x_i$ with $x_3\rightarrow x_3$ and
$\tau\rightarrow \tau$, is a rotation $R(\theta)$ in $SO(2)$ with
$\theta = \pi$; it therefore does not lead to a discrete symmetry.  A
discrete symmetry can be defined by  $x_1\rightarrow x_1$ and
$x_2\rightarrow - x_2$,  as was pointed out in the lattice
studies by Harte and Philipsen~\cite{Philipsen}.  However, the sole
manifestation of this
symmetry is the helicity ``doublets'', $\lambda = \pm J$, for states
with spin $J > 0$. We have already taken this degeneracy into account.

On the other hand, $\tau$-parity remains a discrete symmetry of the
action, as is the case for $QCD_4$. In this paper, we shall define
``parity" for $QCD_3$ as $P\equiv R(\pi)\times P_{\tau}$ where 
$x_i\rightarrow -
x_i$, $x_3\rightarrow x_3$, and $\tau
\rightarrow -\tau$.   This of course is precisely the
parity for the uncompactified 4-d theory with $x_3$ treated as
Euclidean time.

\subsubsection{Graviton, dilaton and axion states}

The graviton $G_{\mu\nu}$ couples as $G_{\mu\nu}T^{\mu\nu}$ as in the
case of $QCD_4$. Because an even number of gluons occur in the field
operators, the charge conjugation for all such states are $C = +$.
For parity, we assume we are in a gauge where the indices of
$G_{\mu\nu}$ do not point along $x_3, r$. From the coupling,
$G_{\mu\nu}\tr[F^{\mu\lambda}F_\lambda^\nu] + \cdots$, we get states
\be
G_{ij}~\rightarrow ~2^{++}\;\; (P_\tau = +),\quad\quad G_{i\tau}~
\rightarrow ~1^{++}\;\; (P_\tau = -),\quad\quad
G_{\tau\tau}\rightarrow ~0^{++}\;\; (P_\tau = +) \; .
\ee
The dilaton couples as $\phi \tr{F^2}$, leading
to
\be
\phi~\rightarrow ~0^{++}\;\; (P_\tau = +),
\ee
and for an axion coupling $C_0\tr(F_{12}F_{\tau 3})$,
\be
C_0 ~\rightarrow ~0^{-+}\;\; (P_\tau = -) \; .
\ee

\subsubsection{Two-form fields and volume scalar}

Consider first the NS-NS 2-form field $B_{\mu\nu}$.  For parity, again
assume that we are in a gauge where the indices of the 2-form do not
point along $x_3, r$. With $i,j=1,2$,
the coupling $B_{ij}\;Sym\tr[F^{ij} W ]\; $
leads to
\be
B_{ij} ~\rightarrow ~0^{+-}\;\; (P_\tau = +) \; ,
\ee
and $B_{i\tau}\;Sym\tr[F^{i\tau} W ]$ to
\be
B_{i\tau} ~\rightarrow ~1^{+-}\;\; (P_\tau = -) \; .
\ee
Finally for the Ramond-Ramond 2-form $C_{\mu\nu}$,
we have the coupling $ C_{12}\; Sym\tr[F_{\tau 3}W]$, so
\be
C_{12} ~\rightarrow ~0^{--}\;\; (P_\tau = -) \; ,
\ee
and $ \epsilon^{ij}C_{\tau i}\;Sym\tr[F_{j3}W]$ giving
\be
C_{i \tau} ~\rightarrow ~1^{--}\;\; (P_\tau = +) \; .
\ee

Finally, as in the case of $QCD_4$, the volume scalar couples as
$h^{\alpha}_{\alpha}\tr
F^4+\cdots$ so that
\be
h^\alpha_\alpha ~\rightarrow ~0^{++}\;\; (P_\tau = +) \; .
\ee
The complete parity and charge conjugation assignments are given in
Table~\ref{tab:classIIB}.

\section{Discussion}

Lastly we turn to the question of how well the strong coupling limit for
the Maldacena dual theory of QCD represents the infrared physics probed
by the glueball spectra. Happily we now have a rather definitive
lattice glueball spectrum by Morningstar and Pearson~\cite{MP} with which to
make comparisons (See right side of Fig.~\ref{fig:comparison} below).

\subsection{Comparison with Lattice Glueball Spectrum}

Originally, claims were made about accurate comparisons to a few
percent for isolated (scalar) mass ratios.  As we pointed out in
Ref.~\cite{bmt1} for $QCD_3$, the lowest mass scalar comes from the
gravitational multiplet, not the dilaton.  A similar spectrum is
observed for $QCD_4$. Consequently such accurate mass ratios were a
misconception. This should not be regarded as a failure, since any
reasonable expectation of a strong coupling approximation should not
give quantitative results. On the other hand, there is a rather
remarkable correspondence of the overall mass and spin structure
between our strong coupling glueball spectrum and the lattice results
at weak coupling for $QCD_4$ (see Fig.~\ref{fig:comparison} below.)
Apparently the spin structure of type IIA supergravity does resemble
the low mass glueball spin splitting.  The correspondence is sufficient
to suggest that the Maldacena duality conjecture may well be correct
and that further efforts to go beyond strong coupling are worthy of
sustained effort.
\begin{figure}[h]
\mbox{\epsfxsize=60mm\epsfysize= 95mm\epsfbox{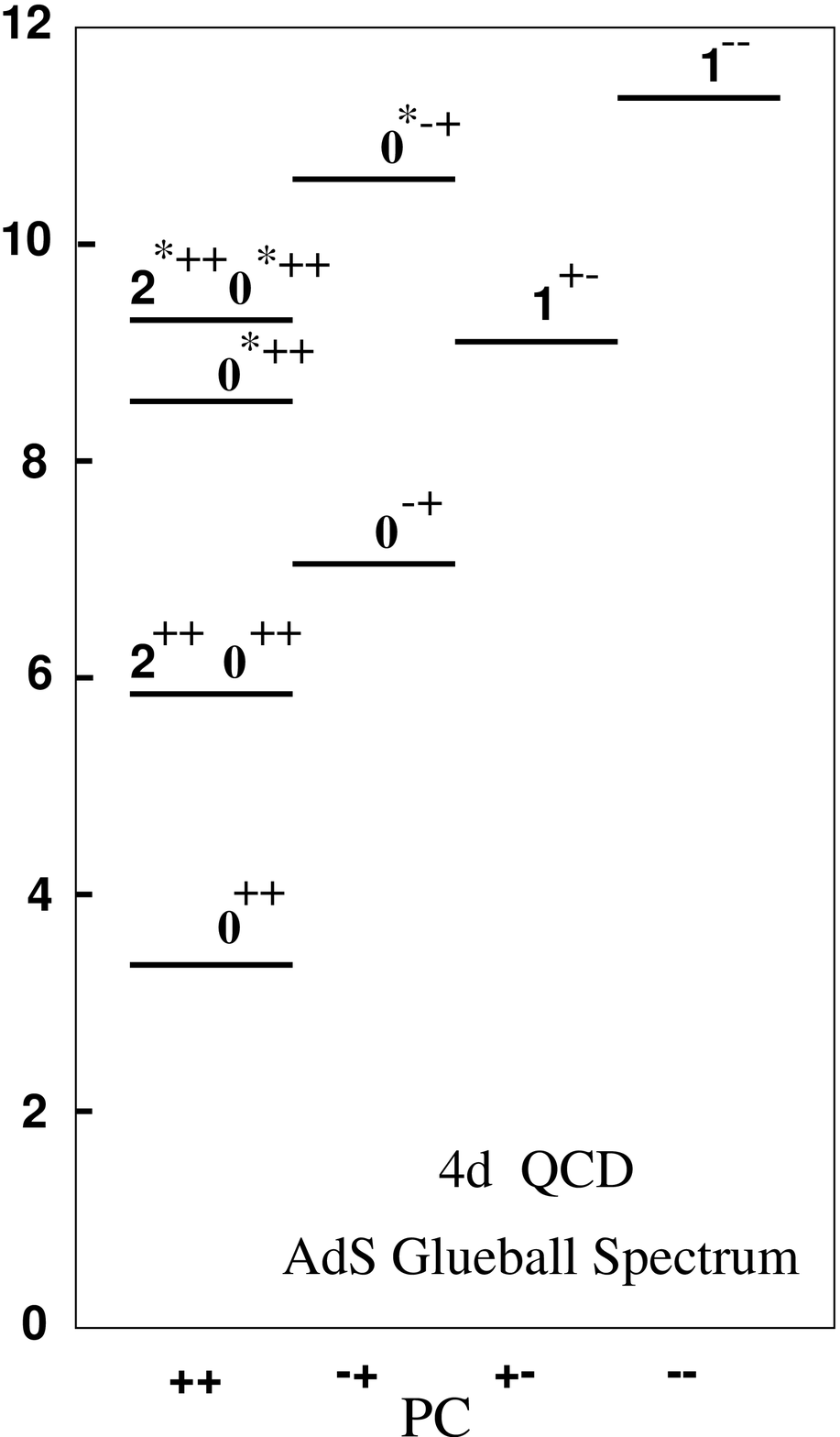}\rule{15mm}{0mm}
        \epsfxsize=75mm\epsfysize= 100mm\epsfbox{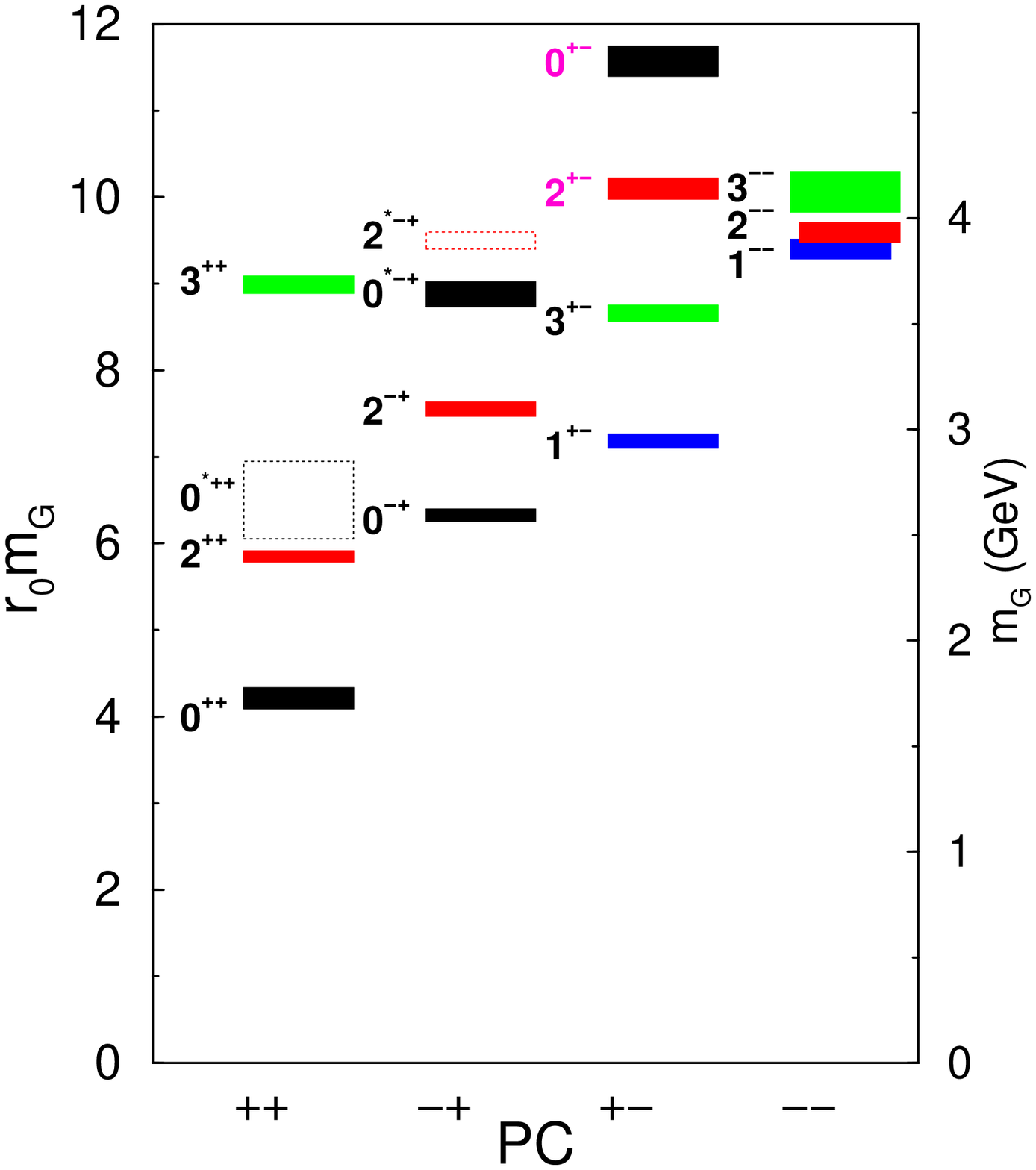}}
        \caption{The AdS glueball spectrum for $QCD_4$ in strong
        coupling (left) compared with the lattice spectrum~\cite{MP} for pure
        SU(3) QCD (right). The AdS cut-off scale is adjusted to set the
        lowest $2^{++}$ tensor state to the lattice results in units of
        the hadronic scale $1/r_0 = 410$ Mev.}  \label{fig:comparison}
\end{figure}

Note that for each value of $PC = (++,-+,+-,--)$, the lowest state is
present in approximately the right mass range.  In addition, the exact
$2^{++}/0^{++}$ degeneracy for AdS strong coupling corresponds to a
relatively small splitting in the lattice calculations. Finally, there
is a radial excitation of the pseudoscalar $0^{*-+}$ that suggests
that even this effect is approximated. This is intriguing because in
the supergravity description the radial mode is a standing wave in the
extra 5th dimension whereas in the lattice it is a conventional
radial mode.  Apparently scale changes in 4-d are being represented by
the distance into the extra ``warped'' 5th axis.

At higher masses the discrepancies increase. One reason is the
obvious fact that on the supergravity side all orbital excitations of
higher spin states are pushed to infinity in strong coupling by virtue
of the divergent string tension,
\begin{equation}
\sigma \equiv \frac{1}{ 2 \pi \alpha'} =   \frac{16 \pi g^2 N}{ 27} \; [\; 1 +
0(\frac{1}{g^2 N})\; ] \;.
\end{equation}
For example, the $3^{++}$ state is a purely stringy effect outside of the
classical limit of supergravity.

Finally, we must emphasize that our comparison is premised on the
neglect of many ``spurious'' states in the strong coupling limit that
are in the wrong superselection sector to survive in the conjectured
weak coupling limit of QCD. For example, all the Kaluza-Klein modes in
the compact thermal $S^1$ manifold and the sphere $S^4$ (or $S^5$)
have masses at the cut-off scale.  (The first mode on the thermal
circle has a KK mass scale $m^2_{KK} = 4$ in type IIB and $m^2_{KK} =
9$ in type IIA in the  units used in Table~\ref{tab:mass3} and
\ref{tab:mass4} respectively.) But these spurious KK modes all carry conserved
$U(1)$ or $R$ charges that are absent in the target theory. We assume
they will disappear in the continuum limit. A subtler situation occurs
in the $QCD_4$ example. Because normal modes in the extreme strong
coupling limit do not distinguish between the compact 11th dimension
and the spatial co-ordinates $x_1,x_2,x_3$ on the brane, the spectrum
actually has an exact $SO(4)$ symmetry. Thus there are additional
states (listed in Table~\ref{tab:classIIA} but ignored in
Fig.~\ref{fig:comparison}) exactly degenerate with the physically
reasonable $2^{++}/0^{++},\; 0^{-+}, \; 1^{+-}$ and $1^{--}$
states. They are all odd under the discrete symmetry of reflecting the
thermal circle (i.e. $P_\tau= -$) so they also lie in another
superselection sector.  A major challenge is to understand how this
$SO(4)$ symmetry is lifted and if the unwanted states remain at the
cut-off in weak coupling. Physically the 11th axis is very
different. The membranes of 11-d M-theory wraps this axis. Another
possibility worth exploring is modifying the background metric with an
orbifold that projects directly onto the even $\tau-$~parity sector
for QCD.

\subsection{Constituent Gluon Picture}

The basic idea behind the AdS/CFT correspondence in the context of
glueballs is similar to an observation made much earlier by Fritzsch
and Minkowski~\cite{FM}, by Bjorken~\cite{BJ} and by Jaffe, Johnson
and Ryzak~\cite{JJR}. Namely that the low mass glueball spectrum
can be qualitatively understood in terms of local gluon interpolating
operators of minimal dimension.

For example, Ref.~\cite{JJR} lists all gauge invariant operators for
dimension $d = 4, 5$ and $6$. Eliminating operators
that are zero by the classical equation of motion and states that
decouple because of the conservation of the energy momentum tensor,
the operators are in rough correspondence with all the low mass
glueballs states, as computed in a constituent gluon~\cite{KS} or bag
model. Indeed more recently Kuti~\cite{Kuti} has pointed out that a
more careful use of the spherical cavity approximation even gives a
rather good quantitative match to the lowest 11 states in the lattice
spectrum.

Consequently it is interesting to compare this set of operators
with the supergravity model. We list below all the operators for $d
\le 6$ except the operators with explicit derivatives (e.g.  $Tr[F D
F]$) and $Tr[F DD F]$ ):

\begin{center}
\begin{tabular}{|l|r|l|r|}
\hline
Dimension     & State        & Operator            &  Supergravity \\
\hline
d = 4      & $0^{++}$             & $Tr(FF) = \vec E^a\cdot \vec E^a - \vec
B^a\cdot \vec B^a$   & $\phi$ \\
d = 4    &  $2^{++}$               &$T_{ij} =   E^a_i\cdot  E^a_j + B^a_i
\cdot   B^a_j - \mbox{trace}$  & $G_{ij}$ \\
d = 4    &  $0^{-+}$             & $Tr(F\tilde F) = \vec E^a\cdot \vec B^a $ &
$C_\tau$ \\
d = 4      & $0^{++}$             & $ 2 \; T_{00} =  \vec E^a\cdot
\vec E^a + \vec
B^a\cdot
\vec B^a$   & $  G_{\tau\tau}$ \\
d = 4    &  $2^{-+}$               & $E^a_i\cdot  B^a_j + B^a_i \cdot  E^a_j
-  \mbox{trace}$  & Absent \\
d = 4    &  $2^{++}$               &$ E^a_i\cdot  E^a_j - B^a_i \cdot  B^a_j
-  \mbox{trace}$  & Absent \\
d = 6    &  $(1,2,3)^{\pm-}$     & $Tr(F_{\mu\nu}
\{F_{\rho\sigma},F_{\lambda\eta}\})
\sim d^{abc}F^a F^b F^c $ & $B_{ij}, C_{ij\tau}$ \\
d = 6    &  $(0,1,2)^{\pm+}$               & $Tr( F_{\mu\nu}
[F_{\rho\sigma},F_{\lambda\eta}]) \sim f^{abc}F^a F^b F^c $  & Absent \\
\hline
\end{tabular}
\end{center}
In this table we have used a Minkowski metric. The classification is
parallel to our discussion of couplings in the Born-Infeld action for
D4 branes (see Sec. 4.1), except that all components and derivatives
in the 5th, i.e., $\tau$, direction are zero and therefore all $P_\tau
= -1$ states are absent (See Table~\ref{tab:classIIA}). The column
on the right lists the supergravity mode that couples (after dropping
$\tau$ components) to each operator.

Several observations are in order.  For the d = 4 operators, there is
complete agreement on the 3 lowest quantum numbers: $J^{PC} =
0^{++},2^{++},0^{-+}$. Also there is agreement on the absence of a low
mass $1^{-+}$ state that Ref.~\cite{JJR} attributes to the
conservation law that decouples the operator corresponding to the
momentum tensor $T_{0i} = \vec E^a \times \vec B^a$.  In this context,
it is worth commenting on the two independent sources for $0^{++}$
states --- the condensate, $\tr (FF)$, and the energy density, $T_{00}
= \half(\vec E^a\cdot \vec E^a + \vec B^a\cdot \vec B^a)$.  Naively
one might conclude the second one coming from the conserved energy
momentum tensor should be dropped, for the same reason we dropped the
operator for momentum conservation, $T_{0i}$.  However in fact for QCD
because of the conformal anomaly for the trace of the energy momentum
tensor it is easy to show that the decoupling argument fails.  
A bag-like model circumvents the decoupling because the bag
itself implies a scale breaking ``vacuum" (empty bag) thus introducing
an extra four-vector, the bag velocity $u_{\mu}$.  Our AdS black hole
background, which is a key ingredient in our approach, also breaks
conformal invariance. Consequently all agree that there is an extra
low mass $0^{++}$ state in addition to the one which in our case is
 degenerate with the tensor $2^{++}$. The lattice data clearly favors 
 this  low mass  $0^{++}$ state, in agreement with our AdS
spectrum. Finally, one rather low mass state, the $2^{-+}$, is missing
in the AdS spectrum.  This state is clearly present in the lattice
spectrum and is identified in the bag model.

At d = 6, we have two states identified in the C = -1 symmetric trace,
$Tr(F_{\mu\nu} \{F_{\rho\sigma},F_{\lambda\eta}\})$: the $1^{+-}$
state for operator $d^{abc} \vec B_a (\vec E_b \cdot \vec E_c)$ and
the $1^{--}$ state for the operator $d^{abc} \vec E_a (\vec E_b \cdot
\vec E_c)$. These states are clearly related to the field content of a
IIA supersymmetric multiplet. The higher spin representation are not
present at strong coupling.  Moreover, we have no states corresponding
to the antisymmetric trace operators, $Tr(F[F,F])$.  It appears that
in the limit where we restrict to supergravity modes and ignore the
massive stringy states, we will be able to obtain only the glueball
state with the symmetric $d^{abc}$ coupling between the group indices,
and not the state with the antisymmetric $f^{abc}$ coupling. If we
consider the chiral primaries of the 4-d Yang-Mills theory (which was
the theory on the boundary before we took the $\tau$ direction to be
compact), then we find that these had the form $Sym~\tr (X^iX^jX^k)$ -
i.e, we have a symmetric trace over the fields.  Other operators that
couple to the supergravity fields will be supersymmetry descendents of
these chiral primaries, but the symmetry in the trace would be
maintained\footnote{We thank W. Taylor for a discussion on this
point.}.  This fact may be related to the observation of
Tseytlin~\cite{Tseytlin} that generalizing the Born-Infeld action to a
non-abelian case gives rise to symmetric trace operators in the field
theory.

One legitimate point of view is simply to suppose that all missing
states must by definition be stringy effects that will be restored in
weak coupling.  However we prefer to look on this as a possible clue
on constructing a better initial geometry for the supergravity/QCD
duality proposals.

\subsection{Strong coupling Expansion for Pomeron Intercept}

We shall end this discussion with a comment on the slope of the leading
glueball trajectory as way to estimate the crossover value for the
bare coupling, where continuum physics might begin to hold.
The Pomeron is the leading Regge trajectory passing through the
lightest glueball state with $J^{PC}=2^{++}$. In a linear
approximation, it can be parameterized by
\begin{equation}
\alpha_P(t) = 2 + {\alpha'_P} (t-m_T^2),
\end{equation}
where we can use the strong coupling estimate
for the lightest tensor mass\footnote{ We have adopted
the normalization in the $AdS$-black hole metric to simplify the
coefficients, e.g., for $AdS^7$, ${\bar g}_{\tau\tau}= r^2-
r^{-4}$. This corresponds to fixing the ``thermal-radius" $R_1=1/3$ so
that $\beta=2\pi R_1 =2\pi/3$.},
\begin{equation}
m_T \simeq [9.86 + 0( \frac{1}{g^2 N} )] \; \beta^{-1} \; .
\end{equation}
Moreover if we make the standard assumption that the closed string tension is
twice that between two static quark sources~\cite{brandhuber}, we also have
a strong coupling expression for the  Pomeron slope,
\begin{equation}
\alpha_P'\simeq [ {27\over 32 \pi g^2 N}+ 0(\frac{1}{g^4 N^2})] \; { \beta^2}.
\end{equation}
Putting these together,
we obtain a strong coupling expansion for the Pomeron intercept,
\begin{equation}
\alpha_P(0) \simeq 2- 0.66 \; (\frac{4 \pi}{ g^2 N}) + 0(\frac{1}{g^4
N^2}) \; .
\end{equation}

Turning this argument around, we can estimate a crossover value between the
strong and weak coupling regimes by fixing $\alpha_P(0)
\simeq 1.2$ at its phenomenological value~\cite{tan0}.  In fact this
yields for $QCD_4$ at $N = 3$ a reasonable value for $\alpha_{strong}
= g^2/4 \pi = 0.176$ for the  crossover. Much more experience with this new
approach to strong coupling must be gained before such numerology can
be taken seriously. However, similar crude argument have proven to be
a useful guide in the crossover regime of lattice QCD.  One might
even follow the general strategy used in the lattice cut-off
formulations. Postpone the difficult question of analytically solving
the QCD string to find the true UV fixed point.  Instead work at a
fixed but physically reasonable cut-off scale (or bare coupling) to
calculate the spectrum. If one is near enough to the fixed point, mass
ratios should be reliable. After all, the real benefit of a weak/strong
duality is to use each method in the domain where it provides the
natural language.  On the other hand, clearly from a fundamental point
of view, finding analytical tools to understand the renormalized
trajectory and prove asymptotic scaling within the context of the
gauge invariant QCD string would also be a major achievement --- an
achievement that presumably would include a proof of confinement
itself.

{\bf Acknowledgments:} We would like to acknowledge useful
conversations with R. Jaffe, A. Jevicki, D. Lowe, J. Kuti, J. Minahan, J. M.
Maldacena, H.  Ooguri, W. Taylor and U-J Wiese.

\def\theequation{A.\arabic{equation}}
\setcounter{equation}{0}

\appendix

\newpage\section{Wave Equations}

In this appendix we outline the derivation of the wave equations that
were used to find the energy levels in the supergravity theory. First
we take the case of $QCD_3$, for which we have the metric,
\be
ds^2 = (r^2-{1\over r^2}) d\tau^2 + r^2 \sum_{i=1,2,3} d x_i^2 +
(r^2-{1\over r^2})^{-1} d r^2 + d\Omega_5^2 \; ,
\ee
where $x_3$ is the Euclidean time direction.

The simplest equation is the scalar wave equation for the dilaton and the
axion. At the linear perturbation level, both satisfy
\be
\phi_{,\mu}{}^{;\mu}={1\over
\sqrt{-g}}[\;\phi_{,\mu}g^{\mu\nu}\sqrt{-g}\;]_{,\nu}=0\;.
\ee

We introduce a  plane wave ansatz,
\be
\phi=T_3(r)e^{ik_3x_3}\; ,
\ee
with zero momentum and mass, $m = i k_3$ providing the equation (8) for  $T_3$
in the text.

Fluctuations in the volume of the sphere $S^5$, which is a fiber at
each point of space time, provides another scalar mode.  This scalar
has an AdS mass squared equal to $32$.  The field equation is
\be
{1\over \sqrt{-g}}[\;\phi_{,\mu}g^{\mu\nu}\sqrt{-g}\;]_{,\nu}-32 \; \phi=0\;,
\ee
$\phi=L_3(r)e^{ik_3x_3}\; $, which reduces to the equation (8) for
$L_3$ in the text.

Next let us address the case of the two-form fields,
$B_{\mu\nu}$ and $C_{\mu\nu}$, which at
the linear level are conveniently combined into
a single complex field, $\tilde B_{\mu\nu} =B_{\mu\nu} + i
C_{\mu\nu}$.  It was shown in
\cite{krv} that this field satisfies the equation,
\be
({\rm
Max}-k(k+4))\tilde
B_{\mu\nu}+2i\epsilon_{\mu\nu}{}^{\rho\lambda\sigma}\partial_\rho
\tilde B_{\lambda\sigma}=0\; ,
\ee
for  $k=0,1,\cdots$ harmonics on $S^5$.
The Maxwell operator is defined by
\be
{\rm Max}\tilde B_{\mu\nu}=H_{\mu\nu\lambda}{}^{;\lambda}\; .
\ee
in terms of the field strength,
\be
H_{\mu\nu\lambda}=\partial_{\mu}\tilde B_{\nu\lambda}+\partial_\nu
\tilde B_{\lambda\mu}
+\partial_\lambda \tilde B_{\mu\nu}\; .
\ee
Since the second order differential operator
factorizes into two first order operators,
solutions fall into two classes,
\be
(2kI+i*D)\tilde B_{\mu\nu}^{(1)}=0, ~~~~(2(k+4)I-i*D)\tilde
B_{\mu\nu}^{(2)}= 0\; ,
\ee
where $(*D B)_{\mu\nu}=\epsilon_{\mu\nu}{}^{\rho\lambda\sigma}\partial_\rho
\tilde B_{\lambda\sigma}$ and $I$ is the identity matrix.

It is convenient to iterate these first order equations to get the
second order equations,
\be
({\rm Max} -k^2)\tilde B_{\mu\nu}^{(1)}=0
, ~~~~({\rm Max} -(k+4)^2)\tilde B_{\mu\nu}^{(2)}=0\; .
\ee
We are interested in fields with no dependence on the coordinates of
the sphere, so we can take $k=0$. It can be shown that the first class
of solutions, $\tilde B_{\mu\nu}^{(1)}$, are pure gauge, so we are only
interested in the second class, which has an effective mass squared of
$16$ for the field $\tilde B_{\mu\nu}^{(2)}$.

As explained in the text, one must of course check that solutions to
the second order equation for $\tilde B_{\mu\nu}^{(2)}$, actually are
valid solution to the original wave equation. This reduce the number of
independent tensor fields, $\tilde B_{\mu\nu}$, from 6 to 3. For
example with the ansatz,
\be
\tilde B_{12}=N_3(r)r^2 e^{ik_3x_3}
\ee
the first order equations will determine $\tilde B_{\tau 3}$ and 
$\tilde B_{r\tau}$
once we have a solution of the second order equation for
$\tilde B_{12}$. This does not place any constraints on the solution for
$\tilde B_{12}$ itself.  We have defined $\tilde B_{12}$ in terms of 
the normalized
coefficient $N_3(r)$ to obtain a hermitian field equation for $N_3$
similar to our earlier scalar mode $T_3$. In a similar manner we can
find the equation for $\tilde B_{1\tau}$. We can solve the second order
equation for this fluctuation without constraint, and then the
requirement arising from the associated first order equations
determines corresponding values of $\tilde B_{23}$ and $\tilde B_{2r}$. After
adopting the ansatz indicated in the main text, we again obtain a
hermitian equation for $M_3(r)$.

The graviton perturbations arise from the Einstein action with a
cosmological constant,
expanded around the given background. We write $G_{\mu\nu}=\bar
g_{\mu\nu}+h_{\mu\nu}$.  the equation for the perturbation
$h_{\mu\nu}$ is
\be
   -{1\over 2}h_{\mu\nu;\lambda}{}^{;\lambda}-{1\over
2}h^\lambda_{\lambda;\mu\nu}+{1\over 2}h_{\mu\lambda;\nu}{}^{;\lambda}+{1\over
2}h_{\nu\lambda;\mu}{}^{;\lambda}+4h_{\mu\nu}=0
\ee
Near the boundary at $r=\infty$, we can choose a gauge to make the
perturbations transverse to $r,x_3$ and traceless. It turns out that
we can maintain this condition for all $r$ for perturbations of the
form $h_{12}$, and for perturbations of the form $h_{1\tau}$. In these
cases the above equation for $h_{\mu\nu}$ gives immediately the wave
equations to be solved. But keeping in mind the decomposition in spin
eigenstates in the $x_1-x_2$ plane, we also find that we have to
consider a spin-0 perturbation which at infinity has the form
$h_{\tau\tau}$, with $h_{11}=h_{22}=-{1\over 2}h_{\tau\tau}$. In this
case we can choose a gauge to make $h_{3\mu}=0$, but for finite $r$ we
will find in this gauge that $h_{rr}\ne 0$ and also that the part
transverse to $r, x_3$ is not traceless.  Thus we have to keep
$h_{\tau\tau} , h_{11}=h_{22}, h_{rr}$ as independent coupled
functions in the analysis. It was shown in \cite{bmt1} how these
equations can be reduced to one effective equation which can then be
solved in the same way as the equations for the other fields. A nicer
choice of gauge was used in \cite{cm} which led to an
equivalent but simpler equation. We will use the latter source for the
equations, especially since the results there include all dimensions,
and so can be used for the case of $QCD_4$ as well.

Now we turn to the case of ${\bf AdS^7\times S^4}$, which is very similar.
The metric is
\be
ds^2 = (r^2-{1\over r^4}) d\tau^2 + r^2 \sum_{i=1,2,3,4,11} d x_i^2 +
(r^2-{1\over r^4})^{-1} d r^2 +{1\over 4} d\Omega_4^2 \; ,
\ee
where we have $x_4$ as the time direction and we note that the radius
of the $S^4$ is
half the curvature radius of the AdS space: this will affect the
masses arising from the
deformations of the sphere.

There is a three-form field $A_{\mu\nu\lambda}$ which behaves in a manner
similar to the
two-form field $\tilde B_{\mu\nu}$ discussed above. Its field 
equation can again
be
factorized into two first order equations, which we iterated to second
order equations. One factor at $k=0$ corresponds to pure gauge, while
the other at $k=0$ has the value $m_{AdS}^2=4(k+3)^2=36$. For the
scalar mode due to fluctuations of the volume of the $S^4$, we get a
$m^2_{AdS} = 4\times 18=72$.  For these considerations
we arrive at the wave equations (22)  given in
the text.

Finally for comparison with  P. van
Nieuwenhuizen  Ref \cite{vantwo}, we note that they
have scaled the radius of the $S^4$ to unity, instead of $\half$.
Our choice was made to keep the radius of the $AdS^7$ equal to
unity, to make the comparison between $AdS^5$ and $AdS^7$ more
natural.  This change in metric scales the squared masses by a factor of $4$
relative to Ref.~\cite{vantwo}.

\def\theequation{B.\arabic{equation}}
\setcounter{equation}{0}

\section{WKB and Variational Estimates}

First we change variable from $r$ to $x\equiv r^2$ and express all
twelve equations in the standard Sturm-Liouville form,
$$ \{ -{d\over dx} \tau(x) {d\over dx}  +w(x) \}  \phi_n(x) = {m_n^2}\>\sigma
(x)
    \phi_n(x) \; ,
$$
where $\tau(x)$, $w(x)$, and $\sigma(x)$ are generalized ``tension",
``external force", and ``mass-density" respectively. In our case, we
have for $QCD_4$: $\tau_4(x) = (x^4-x)$ and  $\sigma_4(x)={x\over 4}$;
for $QCD_3$: $\tau_3(x) = (x^3-x)$ and $\sigma_3(x)= {1\over 4}$.  Force
densities $w(x)$ for all 12 cases are listed in
Table~\ref{tab:variational}.  We shall use this as our starting point
for carrying out variational and WKB analyses.

\subsection{Variational Estimates for $m_0^2$:}
Solving for eigenstates, $\{\phi_n(x)\}$ and their
corresponding eigenvalues, $\{m_n^2\}$, is equivalent to finding
stationery points of the following functional,
\be
\Gamma[\phi] \equiv {\int_1^{\infty} d x \> [ \tau(x) \phi'(x)^2 + w(x)
\phi(x)^2]\over \int_1^{\infty} dx \> \sigma(x)\phi(x)^2} \; ,
\label{eq:var}
\ee
with $m_n^2=\Gamma[\phi_n]$. In particular, the square of the mass for the
lowest state, $m_0^2=\Gamma[\phi_0]$, is the absolute minimum of
$\Gamma[\phi]$.

To be properly defined as  a Sturm-Liouville problem, it is necessary to
impose boundary conditions: $\tau(x)\phi(x)\phi'(x) \rightarrow 0,
$ for $x\rightarrow 1$ and for $x\rightarrow \infty$, in accord
with the boundary conditions stated earlier for our numerical solutions.
Explicit limiting behaviors for all 12 cases are listed in
Table~\ref{tab:variational}.

Given a trial wave function, $\phi(x)$, Eq.~\ref{eq:var}  provides a
variational upper bound for $m_0^2$.  As we have shown in
Ref. \cite{bmt1}, accurate variational estimates for ground-state
masses can be obtained with minimum efforts. Here, we shall only
attempt to obtain simple estimates by choosing trial functions so that
integrals in the equation above can be evaluated analytically.

The simplest possible trial wave function for each case can be chosen
as a product of $\tau(x)$ and $x^{-1}$, as indicated in
Table~\ref{tab:variational}.  Indeed, our variational approach has
served us well by providing a useful consistency check for our
numerical efforts along the way. These are also summarized in
Table~\ref{tab:variational}.

\begin{table}
\begin{tabular}{|c|l|c|c|r|r|}
\hline
Equation & $w(x)$    & $x\rightarrow \infty$  & $x\rightarrow 1$  &Trial
Function   &Bound
for $m_0^2$
\\
\hline\hline
$T_4$   &   0                           &$x^{-3}$       & 1     &$x^{-3}$
&24.0        \\
\hline
$V_4$   & ${9\over 4} x^{-1} (x^3-1)^{-1}$      & $x^{-3}$   &$(x-1)^{1/2}$
&$(x^3-1)^{1/2} x^{-9/2}$
&37.3     \\
\hline
$S_4$   & $-108x^2 (5 x^3-2)^{-2}$  & $x^{-3}$ & 1   &$x^{-3}$   &  8.16 \\
\hline
$N_4$ & $ {9\over 4} (3x^3-1) x^{-1}$ &  $x^{-9/2} $               &1     &
$x^{-9/2} $ &
56.0
\\
\hline
$M_4$ &${9\over 4}x^2(3x^3-2)( x^3-1)^{-1}$ &$x^{-9/2}$
&$(x-1)^{1/2} $ &
$(x^3-1)^{1/2}x^{-6}$     &  93.3 \\
\hline
$L_4$ &   $18 x^2 $   &$ x^{-6}$     & 1   &$ x^{-6}$    &       120     \\
\hline \hline
    $T_3$ &   0    &$x^{-2}$                & 1     &$x^{-2}$   &   12.0    \\
\hline
$V_3 $ & $  x^{-1} (x^2-1)^{-1} $ &$x^{-2}$   &$(x-1)^{1/2}$      &
$(x^2-1)^{1/2} x^{-3} $
&             20.0 \\
\hline
$S_3$    & $ -16 x(3x^2-1)^{-2}   $         &$x^{-2}$              &1
&$x^{-2} $   &   5.55 \\
\hline
$N_3$   &   $ (3x^2-1)  x^{-1}$      &$x^{-2} $            &1  &$x^{-2}$    &
27.0    \\
\hline
$M_3$ & $ x(3x^2-2) (x^2-1)^{-1} $ &$x^{-3}$        &$(x-1)^{1/2}$          &
$(x^2-1)^{1/2}x^{-4}$    & 46.7   \\
\hline
$L_3$ &         $8 x$  & $x^{-4}$   & 1         & $x^{-4}$  &   56.0     \\
\hline
\end{tabular}

\caption{Variational Estimates for $QCD_4$ and $QCD_3$ }
\label{tab:variational}
\end{table}

\subsection{WKB}

As explained in Ref.\cite{bmt1}, we begin a WKB analysis by first bringing our
differential equations from the Sturm-Liouville form into a
radial-Schroedinger form ,
\be
\big ( -{d^2\over d x^2} + V(x;m^2)\big ) \psi(x) = E \psi(x),
\ee
by scaling $\psi(x) \equiv \sqrt {\tau(x)}\phi(x)$. Eigenvalues
$m_n^2$ are found by solving for zero-energy bound states from below,
$E\rightarrow 0^{-}$. The potential $V(x;m^2)$ is given by
\be
V(x;m^2) = \frac{-m^2\sigma(x) + w(x)}{\tau(x)} + \smfrac{1}{2}
({\tau \over \tau}) - \smfrac{1}{4} ({\tau'\over \tau})^{2}.
\ee
To obtain the desired WKB estimate, we simply need to evaluate in the
large $m^2$ limit the following integral: $(n+1/2)\pi =
\int_1^{\infty} d x \sqrt {-\tilde V(x; m^2)}, $ {\it i.e.}, we seek a
WKB condition in the form, $(n+{1\over 2})\pi = s_0 m + s_1 +
0({1\over m}), $ where coefficients $s_0$ and $s_1$ can be explicitly
evaluated.  As explained in Ref. \cite{bmt1}, the ``effective
potential", $\tilde V(x,m^2)$, is $\tilde V(x;m^2) = V(x;m^2) +
{1/( 4 (x-1)^2)}. $

To isolate the $m^2$-dependence, let us write $
\tilde V(x, m^2) \equiv m^2 V_0(x) + \tilde V_1(x)$,
where $V_0(x)= -\sigma(x) /\tau (x),$ i.e., $ V_0^{(4)} (x)= - 1/(4 (x^3-1))$,
and $ V_0^{(3)}(x) = - 1/( 4 x (x^2-1))$. For
instance, one immediately finds that $s_0= \int_1^{\infty} dx \sqrt
{-V_0(x)} $. For $QCD_4$ and $QCD_3$, they are $s_0^{(4)} =
\frac{\sqrt{\pi}}{ 6 } \Gamma({1\over 6})/ \Gamma({2\over 3}) $, and
$ s_0^{(3)}
= \frac{\sqrt{\pi}}{ 4 }
\Gamma({1\over 4})/\Gamma({3\over 4}) $. The remaining piece, $\tilde V_1(x)$,
is listed in Table~\ref{tab:WKB}.

To find coefficient $s_1$, we need to know the behavior of the ratio
of $V_0(x)$ to $\tilde V_1(x)$ near $x\rightarrow 1$ and $x\rightarrow
\infty$.  The dominant behavior of $\tilde V_1(x)$, in these limits,
can be characterized by two indices,
\bea
\tilde V_1(x) \sim {\delta_l^2\over 4 (x-1)^2} + 0 ({1\over (x-1)}),\quad &for&
\quad x\rightarrow 1,\\
    \tilde V_1(x) \sim {\delta^2_r\over 4 x^2} + 0({1\over x^3}),
\quad\quad &for&
\quad x\rightarrow
\infty.
\eea
One finds that $s_1 =-( {\pi\over 2})(\delta_l + \delta _r),$ and $m^2 =
( {\pi\over s_0})^2\{ n^2 + \delta\> n + \>  \gamma\} + 0(n^{-1})\> $
where
\be
\delta = 1+ \delta_l + \delta_r,
\ee
as exhibited in  Table~\ref{tab:WKB}. The values for $\gamma$ are 
found by a fit
to the numerical values of $m_n^2$.

\begin{table}

\begin{tabular*}{150mm}{@{\extracolsep\fill}|c|l|c|c|r|r|}
\hline
Equation \#  &Effective Potential $\tilde V_1(x)$ & $\delta_l$     &$
\delta_r $
&$\delta$                                & $\gamma$          \\
\hline
$T_4$       &${1\over 4 (x-1)^2} + { 2\over x^2}- {9\over 4x^2(x^3-1)^2 } $
&0                   & 3
                 &4
&
3.31                    \\
\hline
$V_4$&${1\over 4 (x-1)^2} + { 2\over x^2} $
    &1                             &3           &5  &   4.90   \\
\hline
$S_4$                                                   &${1\over 4 (x-1)^2}
+ { 2\over x^2} -
{9\over 4x^2(x^3-1)^2 }-
{108x\over(x^3-1) (5 x^3-2)^2 }$
                                      & 0     &3     &4   &     2.20

    \\
\hline
$N_4    $       &${1\over 4 (x-1)^2} + { 2\over x^2}- {9\over 4x^2(x^3-1)^2 }+
{9(3x^3 - 1)\over 4x^2(x^3-1)}$
       &0              &6           &7    &         8.50            \\
\hline
$M_4$&${1\over 4 (x-1)^2}+ { 2\over x^2}+{9(3x^3+1)\over 4x^2(x^3-1)}$
&1       &6              & 8     & 12.5  \\
\hline
$L_4$           &       ${1\over 4 (x-1)^2}+ { 2\over
x^2}- {9\over 4x^2(x^3-1)^2 }+ {18x \over x^3-1}$    &0      &9
&$10$         &
   17.5   \\
\hline\hline
$T_3$&${1\over 4 (x-1)^2}+ { 3 \over 4x^2} - {1\over  x^2(x^2-1)^2} $

&0                     &2               &3
&                    2.01                  \\
\hline
$V_3 $ &${1\over 4 (x-1)^2} + { 3 \over 4x^2}  $
&1                              &2           &4   &          3.28

    \\
\hline
$S_3    $       &${1\over 4 (x-1)^2}+ { 3 \over 4x^2}- {1\over  x^2(x^2-1)^2}
   -{16\over
(x^2-1)(3x^2-1)^2}$
           &0             &2
&3
&      -3.70                 \\
\hline
$N_3$   &${1\over 4 (x-1)^2} + { 3 \over 4x^2}- {1\over  x^2(x^2-1)^2}
+{3x^2-1\over
x^2(x^2-1)}$   &0              &4             &5          &     4.71

              \\
\hline
$M_3$&${1\over 4 (x-1)^2} + { 3 \over 4x^2} + {3x+1\over
x^2(x+1)^2(x-1)}$ &1
      &4               &6    &  7.66     \\
\hline
$L_3$                           &${1\over 4 (x-1)^2}+ {
3 \over 4x^2} - {1\over
x^2(x^2-1)^2} +{8\over x^2-1}$       & 0         & 6      & 7       &
9.50
\\
\hline
\end{tabular*}
\caption{WKB indices for $QCD_4$ and $QCD_3$}
\label{tab:WKB}
\end{table}

\end{document}